\documentclass[aps, prd, 10pt, twocolumn, superscriptaddress,noshowpacs, preprintnumbers, longbibliography,nofootinbib,bibnotes,floatfix]{revtex4-2}
\pdfoutput=1 

\usepackage{amsmath}
\usepackage{amsfonts}
\usepackage{amssymb}
\usepackage{mathtools}
\usepackage{graphics}
\usepackage{tabularx}
\usepackage[export]{adjustbox}
\usepackage{multirow}
\usepackage{citesort}
\usepackage{graphicx}
\usepackage{url}
\usepackage{soul}
\usepackage{physics}
\usepackage{bm}
\usepackage[dvipsnames]{xcolor}
\usepackage[utf8]{inputenc}

\newcommand{\apjs}{Astrophys. J. Supp.}

\newcommand{\mnras}{Mon. Not. R. Astron. Soc.}
\newcommand{\apjl}{Astrophys. J. Let.}
\newcommand{\aap}{Astron. Astrophys.}

\newcommand{\nar}{New Astron. Rev.}

\newcommand{\physrep}{Phys. Rev.}

\newcommand{\ssr}{Space. Sci. Reviews.}

\usepackage{tikz,xcolor}
\usepackage{hyperref}

\definecolor{lime}{HTML}{A6CE39}
\DeclareRobustCommand{\orcidicon}{\hspace{-1mm}
 \begin{tikzpicture}
 \draw[lime, fill=lime] (0,0) 
 circle [radius=0.16] 
 node[white] {{\fontfamily{qag}\selectfont \tiny \,ID}};
 \draw[white, fill=white] (-0.0525,0.095) 
 circle [radius=0.007];
 \end{tikzpicture}
 \hspace{-3mm}
}

\foreach \x in {A, ..., Z}{\expandafter\xdef\csname orcid\x\endcsname{\noexpand\href{https://orcid.org/\csname orcidauthor\x\endcsname}
 {\noexpand\orcidicon}}
}

\begin{document}

\title{Magnetized Shocks Mediated by Radiation from Leptonic and Hadronic Processes}

\author{Shunke Ai\orcidA{}}
\email{shunke.ai@nbi.ku.dk}
\author{Irene Tamborra\orcidB{}}%
\email{tamborra@nbi.ku.dk}
\affiliation{Niels Bohr International Academy and DARK, Niels Bohr Institute, University of Copenhagen, Blegdamsvej 17, 2100, Copenhagen, Denmark}%

\date{\today}

\begin{abstract}
Shocks in astrophysical transients are key sites of particle acceleration. If the shock upstream is optically thick, radiation smoothens the velocity discontinuity at the shock (radiation-mediated shocks). However, in mildly magnetized outflows, a collisionless subshock can form, enhancing the efficiency of  particle acceleration. We solve the hydrodynamic equations of a steady-state, radiation-mediated shock together with the radiative transfer equations accounting for electron and proton acceleration. Our goal is to explore the impact of the magnetic field and non-thermal radiation on the shock structure and the resulting spectral distribution of photons. To this purpose, we assume a relativistic upstream fluid velocity ($\Gamma_u = 10$) and investigate shock configurations with variable upstream magnetization ($\sigma_u = 0$, $10^{-8}$, $10^{-4}$, $0.1$, and $0.3$). We find that synchrotron self-absorption alters the shock profile for $\sigma_u \gtrsim 10^{-8}$, with resulting changes  up to $100\%$ in the bulk Lorentz factor at the shock; for $\sigma_u \gtrsim 0.1$, a prominent subshock forms. The spectral energy distributions of upstream- and downstream-going photons are also altered. Radiative processes linked to  accelerated  protons are responsible for a high-energy ($\gtrsim 10$~GeV) tail in the photon spectrum; however, the radiation flux and pressure are negligibly affected with  consequent minor impact on the shock structure. Our work highlights  the importance of  coupling  the shock hydrodynamics to the transport of photons, electrons, protons, and intermediate particles to forecast the multi-messenger emission from astrophysical transients.
\end{abstract}

\maketitle

\section{\label{sec:introduction}Introduction}

Shocks are ubiquitous in a wide range of astrophysical transients. Across the shock plane, the thermodynamic variables characterizing the system exhibit a discontinuity due to the sudden compression or expansion of matter. Particles crossing the shock plane can be accelerated~\cite{1949PhRv...75.1169F,1987PhR...154....1B}, leading to the production of non-thermal photons, cosmic-rays, and high-energy neutrinos~\cite{2025NatRP...7..285T,2020NewAR..8901543M,2017hsn..book..967W,2020RvMP...92d5006V,ParticleDataGroup:2024cfk,2023PhRvD.108h3035G}. 

A special class of shocks, relevant to the early multi-messenger emission from astrophysical transients, is the one of radiation-mediated shocks~\cite{2020PhR...866....1L,1976ApJS...32..233W, 2008PhRvL.100m1101L}. These  are of relevance, e.g., for some supernova types, (low-luminosity) gamma-ray bursts, and neutron-star merger remnants. Understanding the physics of radiation-mediated shocks is crucial, since the particle emission linked to the breakout of such shocks carries important information on e.g.~the explosion mechanism, the properties of the outflow, and the shock structure. Moreover,  observations of gamma-ray bursts challenge the synchrotron origin of the photon spectrum; (sub-)photospheric dissipation has been proposed to  possibly lead to  distortions of the Wien photon distribution~\cite{2005MNRAS.363L..61R,2017IJMPD..2630018P,2017SSRv..207...87B,2024ApJ...961L...7R,2025ApJ...983...34R,2023ApJ...956...42S,2013ApJ...764..143V,2025arXiv251108684A}. But, self-consistent models accounting for the feedback of radiation on the shock structure are yet poorly investigated.


Shocks are mediated by radiation  when a shock travels rapidly through an optically thick plasma. In this case, radiation pressure dominates over the gas thermal pressure in the downstream region.  Upstream photons scatter off the incoming fast material, decelerating it prior to its arrival at the shock front. Due to this pre-deceleration, the discontinuity across the shock front is significantly smoothed. As a consequence, radiation-mediated shocks typically lack efficiency in particle acceleration~\cite{2013PhRvL.111l1102M}. The properties of non-relativistic and relativistic radiation-mediated shocks have been thoroughly investigated analytically and later verified by numerical simulations~\cite{1976ApJS...32..233W,2008PhRvL.100m1101L,2010ApJ...716..781K,2010ApJ...725...63B,2019ApJ...879...83L,2017SSRv..207...87B,2018MNRAS.474.2828I}. In particular, for relativistic and non-magnetized shocks, it was pointed out that the shock structure can be largely affected by pair production processes. The latter lead to copious production of photons, which in turn affect the shock structure and temperature in the downstream region \cite{2010ApJ...716..781K,2010ApJ...725...63B}. 

In magnetized media, the magnetic field component perpendicular to the shock normal would be compressed. If the magnetic pressure in the shock downstream is not negligible, a significant jump across the shock plane lingers--a collisionless subshock forms, embedded in the radiation-mediated shock~\cite{2017SSRv..207...87B}. Although mediated by radiation, particles may be accelerated at the subshock. 

If protons are accelerated at the subshock, hadronic processes such as proton-proton and proton-photon interactions further modify the photon spectrum~\cite{2010PhRvD..82i9901K,2006PhRvD..74c4018K}. Recently, relying on three-dimensional general relativistic magnetohydrodynamic jet simulations, Refs.~\cite{2024ApJ...961L...7R,2025ApJ...983...34R} discussed the possible existence of subshocks in the sub-photospheric region of short gamma-ray bursts. In this case, the photon and  high-energy neutrino signals should be molded by the hadronic cascades occurring below the photosphere. However, it has never been explored whether the  radiation produced in the aftermath of hadronic processes would further exert a feedback effect on the shock structure.

In this work, we consider a static radiation-mediated shock building on the approach presented in Ref.~\cite{2010ApJ...725...63B}. For the first time, we explore  modifications of the shock structure, accounting for the medium magnetization and the feedback of radiation generated by both thermal electrons and high-energy protons. Note that, in Ref.~\cite{2017ApJ...838..125B}, a preliminary investigation on the impact of magnetization was presented, however pair processes were accounted for in an approximated fashion and the impact of radiation on the formation of the subshock for varying $\sigma_u$ was studied semi-analitically; on the other hand, Ref.~\cite{2010ApJ...725...63B} focused on the impact of leptonic processes in non-magnetized shocks.

This paper is organized as follows. We outline the physics of radiation-mediated shocks and radiation transfer in Sec.~\ref{sec:classical}. The numerical setup employed to   simulate the formation of subshocks and the feedback of radiation on the shock structure is presented in Sec.~\ref{sec:numerics}. Section~\ref{sec:RMSleptonic} explores the feedback on the shock structure due to radiation produced by means of leptonic processes in magnetized media. We extend our findings by accounting for the radiation produced by hadronic processes in Sec.~\ref{sec:RMShadronic}. Finally, we conclude and discuss our results in Sec.~\ref{sec:conclusions}. In addition, details on the processes leading to production of radiation through thermal leptons and hadronic processes are provided in Appendices~\ref{app:leptonic} and \ref{app:acc_cool}, respectively.

\section{\label{sec:classical} Model for radiation-mediated shocks}

When a shock occurs in an optically thick region, the photons radiated by leptons in the downstream (e.g.~through Compton scattering, bremsstrahlung, pair annihilation, and synchrotron radiation) contribute to build up radiation pressure and energy flux, thus altering the structure of the shock. In this section, we introduce the hydrodynamic and radiation transfer equations adopted to explore the evolution of the shock structure. To this purpose, we closely follow Ref.~\cite{2010ApJ...725...63B}, albeit extending the  equations to the case of a magnetized shock. We also outline the physics underlying the formation of a subshock and the feedback of radiation on the shock structure. 

We consider a steady-state planar shock for simplicity. Such configuration holds in an optically thick region, where photons are trapped. Once the upstream region becomes optically thin, the velocity jump at the subshock increases, and the shock structure is expected to evolve with time. 

Hereafter, we use the subscript ``sh'' to indicate physical quantities defined in the shock frame; when not otherwise specified, quantities are defined in the fluid comoving frame. The subscript ``u'' represents characteristic quantities in the far upstream, which have not been affected by radiation, while we do not use any suffix for  quantities characterizing any other location (either upstream or downstream). 

\subsection{Shock hydrodynamics and radiation transfer}
\label{sec:dynamics} 

For a fluid consisting of protons and electrons and with an ordered magnetic field, the magnetization parameter is  
\begin{eqnarray}
    \sigma = \frac{B^2}{4\pi n_p m_p c^2}\, ,
\end{eqnarray}
where $n_p$ is the proton number density, $B$ is the magnetic field strength, $m_p$ stands for the proton mass, and $c$ is the speed of light. The maximum fast magnetosonic speed  can be calculated as
\begin{eqnarray}
   v_{\rm F,max} = c \sqrt{\frac{\sigma}{1+\sigma} + \frac{c_s^2}{c^2}\frac{1}{1+\sigma}}\, .
\end{eqnarray}
This corresponds to a Lorentz factor
\begin{eqnarray}
    \Gamma_{\rm F,max} = c  \sqrt{\frac{1+\sigma}{c^2 - c_s^2}}\, ,
\label{eq:FMS}
\end{eqnarray}
with $c_s$ being the local speed of sound ($c_s = 0$ for a cold fluid). In the comoving frame of a magnetized plasma, when a disturbance in the fluid velocity exceeds this critical speed, a magnetohydrodynamic shock wave is excited. In the shock frame, a supersonic flow enters the shock region, undergoes deceleration, and eventually propagates away from the shock at subsonic velocity. 
\begin{figure*}
\centering
\includegraphics[width=1.98\columnwidth]{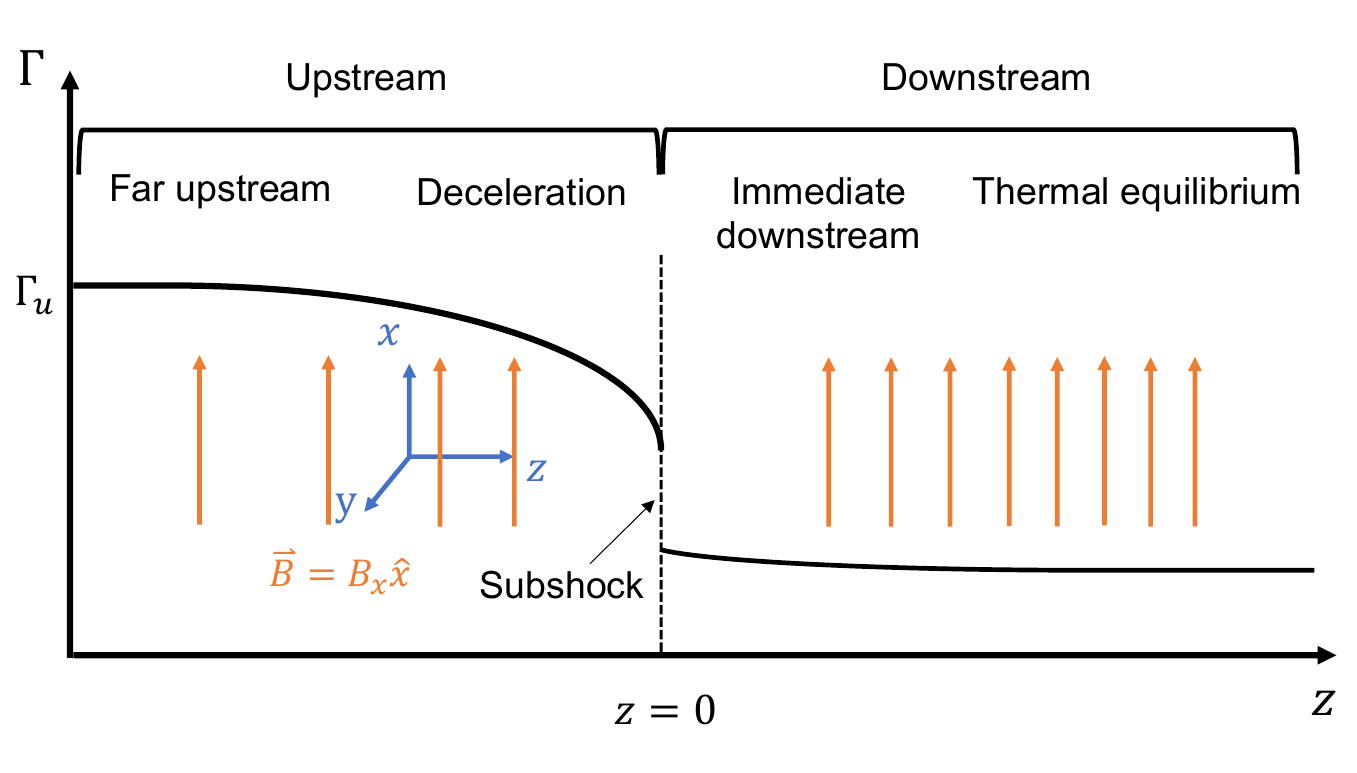}
\caption{Sketch of the bulk Lorentz factor profile of a radiation-mediated magnetohydrodynamic shock along the $z$ direction. The vertical orange arrows represent the magnetic field lines, which are more intense in the downstream region, where the proton density is higher. The magnetic field component parallel to the shock normal does not directly affect the shock dynamics and it is not plotted here for simplicity. We distinguish four regions characterizing the shock structure: the far upstream, the deceleration region, the immediate downstream, and the thermal equilibrium region, from left to right, respectively. A subshock forms at $z=0$ allowing for efficient particle acceleration.}
\label{fig:sketch}
\end{figure*}

As shown in Fig.~\ref{fig:sketch}, we consider a steady-state planar shock with its front perpendicular to  the $z$ direction. The shock moves with velocity $c \beta_u$ through a cold and magnetized plasma made of protons and electrons, with  upstream comoving  density of protons  $n_{p,u}$ and magnetization  $\sigma_u$. The magnetic field is assumed to be perpendicular to the shock normal, along the $x$ axis (the magnetic field component parallel to the shock normal crosses the shock without any compression, hence it does not directly influence the shock dynamics and it is not represented in Fig.~\ref{fig:sketch}; but this component may influence the efficiency of particle acceleration). 

The structure of the radiation-mediated shock can be divided into four regions: far upstream, deceleration region, immediate downstream, and thermal equilibrium region. Photons are initially generated in the immediate downstream because of dissipation of  internal energy deposited at the shock. Then, photons travel towards the upstream and the downstream. The downstream achieves thermal equilibrium through particle scatterings and reprocessing of photons. The photons propagating upstream interact with the fluid flowing towards the shock plane, thereby decelerating it and smoothing out its velocity jump. In the non-magnetized case, the subshock is less prominent~\cite{2010ApJ...725...63B, 2017ApJ...838..125B}. 

To derive the governing equations describing the system dynamics, the following assumptions are made: (1) the perpendicular component of the magnetic field is frozen into protons (i.e., $\sigma \propto B^2/n_p \propto n_p$) and is compressed as the proton number density increases; (2) the electron (positron) fluid and the ion fluid move at the same velocity.

The conservation equations for  energy flux and momentum flux are: 
\begin{eqnarray}
\frac{d}{dz} T_{\rm sh}^{0z} = 0\, ,
\label{eq:energy_cons}
\end{eqnarray}
and 
\begin{eqnarray}
\frac{d}{dz} T_{\rm sh}^{zz} = 0\, ,
\label{eq:momentum_cons}
\end{eqnarray}
where $T_{\rm sh}^{\alpha \beta} = T_{\rm fl,sh}^{\alpha \beta} + T_{\rm rad,sh}^{\alpha \beta} + T_{\rm em,sh}^{\alpha \beta}$ is the total energy-momentum tensor; in $T_{\rm sh}^{\alpha \beta}$, the contributions of the fluid including pairs and proton-electron plasma, radiation, and the electromagnetic field are denoted with the subscripts ``fl,'' ``rad,'' and ``em,'' respectively. 

Assuming that protons, electrons and positrons in the fluid have the same average velocity $\beta = v/c$ (which corresponds to a Lorentz factor $\Gamma = \sqrt{1 - 1/\beta^2}$), $T_{\rm fl, sh}^{\alpha \beta}$ can be expressed as 
\begin{eqnarray}
T_{\rm fl,sh}^{0z} = \Gamma^2 \beta (\rho c^2 +e_{\rm fl} + P_{\rm fl})
\end{eqnarray}
and
\begin{eqnarray}
T_{\rm fl,sh}^{zz} = P_{\rm fl} + \Gamma^2 \beta^2 (\rho c^2 +e_{\rm fl} + P_{\rm fl})\, ,
\end{eqnarray}
where $\rho =  n_p m_p + (n_{e^{+}} + n_{e^{-}}) m_e$, $e_{\rm fl} = {3}/{2}f(T)(n_e+n_+)T$, and $P_{\rm fl} = (n_{e^{+}} + n_{e^{-}})T$ represent the density, internal energy, and thermal pressure, respectively. In the equations above, $m_e$ represents the electron  mass, while the number densities of protons, electrons, and positrons are $n_p$, $n_{e^-}$ and $n_{e^+}$, respectively. The plasma temperature is $T$; the function $f(T)$ approximates the equation of state for a Maxwell-Boltzmann distribution: 
$f(T) = {1}/{2}\ {\rm tanh}[{{\rm ln}(T/m_e c^2)+0.3}/{1.93}] + {3}/{2}$.

The components $T_{\rm sh,rad}^{\alpha \beta}$ of the energy-momentum tensor read as~\footnote{The radiation intensity is a function of angle ($\mu$), frequency ($\nu$), and position ($z$). For simplicity, we use $I_{\nu}(\mu)$ instead of $I(\mu,\nu,z)$ hereafter.}
\begin{eqnarray}
T_{\rm rad, sh}^{0z} = c^{-1} F_{\rm rad, sh} = \iint d\Omega_{\rm sh} \mu_{\rm sh} d\nu_{\rm sh} I_{\nu,{\rm sh}}(\mu_{\rm sh})
\end{eqnarray}
and
\begin{eqnarray}
T_{\rm rad, sh}^{zz} = P_{\rm rad, sh} = c^{-1} \iint d\Omega_{\rm sh} \mu_{\rm sh}^2 d\nu_{\rm sh} I_{\nu,{\rm sh}}(\mu_{\rm sh})\ ,
\end{eqnarray}
with $I_{\nu,{\rm sh}}$ being the radiation intensity and $\nu_{\rm sh}$  the photon frequency; the differential solid angle is  $d\Omega = 2\pi d\mu$, assuming azimuthal symmetry, and $\mu = \cos \theta$. 

The components $T_{\rm sh,em}^{\alpha \beta}$ are
\begin{eqnarray}
T^{0z}_{\rm em,sh} &=& \frac{1}{4\pi} (\vec{E}_{\rm sh}\times \vec{B}_{\rm sh}) 
 =  \Gamma^2 \beta \rho c^2 \sigma\, ,\nonumber \\
T^{zz}_{\rm em,sh} &=& \frac{1}{8\pi}(E_{\rm sh}^2 + B_{\rm sh}^2) 
 =  \frac{1}{2} \Gamma^2 (1+\beta^2) \rho c^2 \sigma\, ,
\end{eqnarray}
where we assume that $\vec{B}_{\rm sh} = B_{\rm x,sh} \hat{x}$, so that $\vec{E}_{\rm sh} = -{\vec{v}}/{c} \times \vec{B} = -\beta B_{\rm x, sh} \hat{y}$ is obtained from Ohm's law. 

In addition to the conservation of energy and momentum fluxes, the conservation of the particle number as well as the magnetic flux should be satisfied:
\begin{eqnarray}
n_p = n_{p,u}  \frac{\Gamma_u \beta_u}{\Gamma \beta}\, , 
\label{eq:p_number_cons}
\end{eqnarray}
\begin{eqnarray}
\frac{d(\Gamma \beta n_+)}{dz} = \frac{Q_{e^{\pm}}}{c}\, ,
\label{eq:e_number_cons}
\end{eqnarray}
\begin{eqnarray}
    \sigma = \sigma_u \frac{\Gamma_u \beta_u}{\Gamma \beta}\, ,
\label{eq:B_flux_cons}
\end{eqnarray}
with $Q_{e^{\pm}}$ being the pair production rate. The radiation transfer equation reads as
\begin{eqnarray}
\mu_{\rm sh}\frac{dI_{\nu_{\rm sh}}(\mu_{\rm sh})}{dz} &=& \eta_{\rm sh}(\mu_{\rm sh},\nu_{\rm sh}) \nonumber \\
&-& I_{\nu_{\rm sh}}(\mu_{\rm sh}) \chi_{\rm sh}(\mu_{\rm sh},\nu_{\rm sh})  
\label{eq:radiation_transfer}
\end{eqnarray}
where $I_{\nu_{\rm sh}}$, $\eta_{\rm sh}$, and $\chi_{\rm sh}$ represent the intensity, emission  and absorption coefficients, in the shock frame, respectively. The photon frequency and the direction along which that  radiation intensity is defined are $\nu_{\rm sh}$ and $\mu_{\rm sh}$, respectively, where $\mu = {\rm cos} \theta$  with $\theta$ representing the angle with respect to the $z$ direction. The radiation processes entering these coefficients are introduced in Sec.~\ref{sec:radiation}.

In order to facilitate our  calculations, we introduce the following dimensionless quantities (valid in any reference frame):
$\hat{T} = {T}/{(m_e c^2)}$, 
$\hat{\nu} = {h\nu}/{(m_e c^2)}$,
$a_{e^{\pm}} = n_{e^{+}} / n_p$, 
$d\hat{\tau} = \Gamma (1+\beta)(n_e + n_+)\sigma_T dz$,
and $\hat{I} = {I}/{\Gamma_u^2 \beta_u n_{p,u} (m_p/m_e)hc}$; $h$ is the Planck constant, and  $\sigma_T$ is the Thomson cross section. Hence, the energy and momentum conservation equations (Eqs.~\ref{eq:energy_cons} and \ref{eq:momentum_cons}) become
\begin{widetext}
\begin{eqnarray}
    \frac{\Gamma}{\Gamma_u}\left\{(1+\sigma) + (1+2a_{e^{\pm}})\frac{m_e}{m_p} \left[1 + \hat{T}(1+\frac{3}{2}f(\hat{T}))\right] \right\}  + 2\pi \hat{F}_{\rm rad,sh} = \left(1+\frac{m_e}{m_p}\right)+\sigma_u\, ,
\label{eq:E_cons_normal}
\end{eqnarray} 
\begin{align}
\frac{\Gamma \beta}{\Gamma_u \beta_u} \left\{(1+\sigma)+\frac{1}{2}\frac{\sigma}{\Gamma^2\beta^2} +(1+2a_{e^{\pm}})\frac{m_e}{m_p} \left[1+\hat{T}\left(1+\frac{3}{2}f(\hat{T})+\frac{1}{\Gamma^2\beta^2}\right)\right]\right\}
& + 2\pi\frac{1}{\beta_u}\hat{P}_{\rm rad,sh} \nonumber
\\[4pt] 
&= \left(1+\frac{m_e}{m_p}\right) +\sigma_u +\frac{1}{2}\frac{\sigma_u}{\Gamma_u^2\beta_u^2}\, .
\label{eq:M_cons_normal}
\end{align}

\end{widetext}
The radiative transfer equation (Eq.~\ref{eq:radiation_transfer}) becomes
\begin{eqnarray}
\mu_{\rm sh}\frac{d\hat{I}_{\hat{\nu}_{\rm sh}}(\mu_{\rm sh})}{d\hat{\tau}} &=& \hat{\eta}_{\rm sh}(\mu_{\rm sh},\hat{\nu}_{\rm sh}) \nonumber \\
&-& \hat{I}_{\hat{\nu}_{\rm sh}}(\mu_{\rm sh}) \hat{\chi}_{\rm sh}(\mu_{\rm sh},\hat{\nu}_{\rm sh})\, .
\label{eq:radiation_transfer_normal}
\end{eqnarray} 
The evolution of the pair density is given by 
\begin{eqnarray}
    \frac{da_{e^{\pm}}}{d\hat{\tau}} = \hat{Q}_{e^{\pm}}\, .
\label{eq:pair_prodution_normal}
\end{eqnarray}
In the  equations above, we have introduced the following  dimensionless quantities: 
$\hat{F}_{\rm rad,sh} = {F_{\rm rad,sh}}/{(2\pi\Gamma_u^2 \beta_u n_{p,u} m_p c^3)}$, 
$\hat{P}_{\rm rad,sh} = {cP_{\rm rad,sh}}/{(2\pi \Gamma_u^2 \beta_u n_{p,u} m_p c^3)}$, 
$\hat{\eta} = {(\eta m_e)}/{[\Gamma (1+\beta)\sigma_T (n_{e^+} + n_{e^-}) m_p \Gamma_u^2 \beta_u n_{p,u} hc]}$, 
$\hat{\chi} = {\chi}/{[\Gamma (1+\beta)\sigma_T(n_{e^+}+n_{e^-})]}$, 
and $\hat{Q}_{e^{\pm}} = {Q_{e^{\pm}}}/{[\Gamma^2 \beta (1+\beta)n_p(n_{e^+}+n_{e^-})\sigma_T c]}$.

\subsection{\label{sec:radiation} Radiation from leptonic processes}
We account for the following leptonic processes producing radiation: Compton scattering, pair production and annihilation, bremsstrahlung and absorption, as well as synchrotron and self-absorption. The emission and absorption coefficients for the first three processes are provided in Ref.~\cite{2010ApJ...725...63B}; a similar procedure is adopted for  synchrotron. In the following, we provide the main expressions for the emission and absorption coefficients and refer the interested reader to Appendix~\ref{app:leptonic} for details.

For Compton scattering, the normalized emission and absorption coefficients are:
\begin{eqnarray}
\hat{\eta}_{s, {\rm sh}}(\hat{\nu}_{\rm sh},\mu_{\rm sh}) &=& \frac{1}{[\Gamma (1 - \beta\mu_{\rm sh})]^2} \times \frac{1}{\Gamma(1+\beta)}  \\
&&\times \iint d\Omega^{\prime}d\hat{\nu}^{\prime} \frac{d\tilde{\sigma_s}}{d\hat{\nu}^{\prime}}(\hat{\nu}^{\prime},\mu^{\prime}\rightarrow \hat{\nu},\mu)\hat{I}_{\hat{\nu}^{\prime}}(\mu^{\prime})\, , \nonumber
\label{eq:eta_s}
\end{eqnarray}
\begin{eqnarray}
\hat{\chi}_{s, {\rm sh}}(\hat{\nu}_{\rm sh},\mu_{\rm sh}) = \Gamma (1 - \beta\mu_{\rm sh}) \times \frac{1}{\Gamma(1+\beta)}\tilde{\sigma}_c(\hat{\nu},\hat{T})\, , \nonumber \\
\label{eq:chi_s}
\end{eqnarray}
where  $\hat{\nu}^{\prime}$ and $\mu^{\prime}$ represent, respectively, the photon frequency and the direction of  radiation intensity of the incoming photons in the fluid comoving frame;   $\tilde{\sigma} = \sigma / \sigma_T$ (with $\sigma_c$ being the  Compton cross section and $\sigma_T$ the Thomson cross section), and $d\tilde{\sigma_s}/d\hat{\nu}^{\prime}$ is the differential cross section of a photon with frequency $\hat{\nu}^{\prime}$  scattered to  $\hat{\nu}$  (cf.~Eq.~\ref{eq:Compton}).  

For pair production and annihilation, the normalized emission and absorption coefficients are:
\begin{eqnarray}
\hat{\eta}_{e^{\pm}, {\rm sh}}(\hat{\nu}_{\rm sh},\mu_{\rm sh}) &=& [\Gamma(1-\beta \mu_{\rm sh})]^{-2}  \\
&\times& \frac{(a_{e^{\pm}} + 1)a_{e^{\pm}} \hat{\nu} f_{e^{\pm}}(\hat{\nu},\hat{T}) r_{e^{\pm}}(\hat{T})}{4\pi (2a_{e^{\pm}} + 1)\Gamma_u \Gamma^2 \beta (1 + \beta)}\frac{m_e}{m_p}\, , \nonumber
\label{eq:eta_pair}
\end{eqnarray}

\begin{eqnarray}
&&\hat{\chi}_{\gamma\gamma, {\rm sh}}(\hat{\nu}_{\rm sh},\mu_{\rm sh}) = \frac{\Gamma_u \beta (m_p/m_e)}{(1+\beta)(2a_{e^{\pm}} + 1)} \nonumber \\ 
&&~~~~~~\times \int \tilde{\sigma}_{\gamma\gamma}(\hat{\nu}_{\rm sh},\hat{\nu}^{\prime}_{\rm sh},\mu_{\rm sh},\mu^{\prime}_{\rm sh}) \frac{\hat{I}_{\nu^{\prime}_{\rm sh}}(\mu^{\prime}_{\rm sh})}{\hat{\nu}^{\prime}_{\rm sh}} (1-{\rm cos}\theta_1) \nonumber \\ 
&&~~~~~~\times~\Theta[\hat{\nu}_{\rm sh}\hat{\nu}^{\prime}_{\rm sh}(1-{\rm cos }\theta_1)-2)]d\Omega^{\prime}_{\rm sh} d\hat{\nu}^{\prime}_{\rm sh}\, ,
\label{eq:chi_pair}   
\end{eqnarray}

with $r_{e^{\pm}}$ standing for the pair annihilation rate, and $f_{e^{\pm}}(\hat{\nu}_{\rm sh},\hat{T})$ describing the spectrum of photons produced by pairs. The cosine of the angle between $\mu_{\rm sh}$ and $\mu^{\prime}_{\rm sh}$ is $\cos\theta_1$;  the average of the latter with respect to the azimuthal angle is  $\langle 1 - \cos\theta_1 \rangle_\phi = 1 - \mu_{\rm sh} \mu^{\prime}_{\rm sh}$. The cross section for pair production is $\sigma_{\gamma \gamma}$. The pair production and annihilation rates  are 
\begin{equation}
\hat{Q}_{e^{\pm}, {\rm prod}} = \frac{\Gamma_u m_p}{2 m_e} \int \frac{\hat{I}_{\hat{\nu}_{\rm sh}}(\mu_{\rm sh})}{\hat{\nu}} \, \hat{\chi}_{\gamma \gamma}(\mu_{\rm sh},\hat{\nu}_{\rm sh}) \, d\hat{\nu}_{\rm sh} \, d\Omega_{\rm sh}\, ,
\label{eq:Q_prod}
\end{equation}
\begin{eqnarray}
\hat{Q}_{e^{\pm}, {\rm ann}} = - \frac{a_{e^{\pm}} (a_{e^{\pm}} + 1)r_{e^{\pm}}(\hat{T})}{2\Gamma^2 \beta(1+\beta)(1+2 a_{e^{\pm}})}\, ,
\label{eq:Q_ann}
\end{eqnarray}
with the net pair production rate being $\hat{Q}_{e^{\pm}} = \hat{Q}_{e^{\pm}, {\rm prod}} + \hat{Q}_{e^{\pm}, {\rm ann}}$. 

The  emission and absorption  coefficients for bremsstrahlung emission and absorption are 
\begin{eqnarray}
&&\hat{\eta}_{\rm ff, sh}(\hat{\nu}_{\rm sh},\mu_{\rm sh}) = [\Gamma(1-\beta \mu_{\rm sh})]^{-2} \\
&&~~~~~~\times\frac{\alpha_e (m_e/m_p)}{\pi^2 \Gamma_u \Gamma^2 (1+\beta)\beta (1+2a_{e^{\pm}})}\sqrt{\frac{2}{\pi}} \frac{e^{-\hat{\nu}/\hat{T}}}{\sqrt{\hat{T}}} \lambda_{\rm ff}\, ,  \nonumber 
\label{eq:eta_ff}
\end{eqnarray}
\begin{eqnarray}
&&\hat{\chi}_{\rm ff, sh}(\hat{\nu}_{\rm sh}, \mu_{\rm sh}) = [\Gamma (1 - \beta \mu_{\rm sh})] \\
&&~~~~~~\times \frac{\alpha_e h^3 \lambda_{\rm ff}}{\sqrt{2} \pi^{5/2} m_e^3 c^3 \Gamma (1+\beta)(1+2a_{e^{\pm}})}\frac{n_p(1-e^{-\hat{\nu}/\hat{T}})}{\hat{\nu}^3 \sqrt{\hat{T}}}\, , \nonumber
\label{eq:chi_ff}
\end{eqnarray}
where $\alpha_e$ is the fine structure constant, $\lambda_{\rm ff} = {\pi}/{(2\sqrt{3})} g_s$, and $g_s$ is the Gaunt factor. 

In the proximity of a shock  with a relativistic upstream flow, downstream electrons and positrons can reach relativistic temperatures. When the fluid is magnetized, synchrotron radiation and self-absorption become unavoidable. The normalized coefficients for synchrotron emission and self-absorption are
\begin{eqnarray}
&&\hat{\eta}_{{\rm syn, sh}}(\hat{\nu}_{\rm sh},\mu_{\rm sh}) = [\Gamma(1-\beta \mu_{\rm sh})]^{-2}  \\
&&~~~~~~\times  \frac{\sqrt{3}e^3 B}{4\pi m_e c^2}\int n_p (1+2x_+)f_{e^{\pm}}(\gamma_e)F\left(\frac{\nu}{\nu_c}\right)d\gamma_e\ , \nonumber
\label{eq:eta_syn}
\end{eqnarray}
\begin{eqnarray}
&&\hat{\chi}_{{\rm syn, sh}}(\hat{\nu}_{\rm sh}, \mu_{\rm sh}) = [\Gamma (1 - \beta \mu_{\rm sh})] \frac{c^2}{8\pi h\nu^3} n_p (1 + 2a_{e^{\pm}}) \nonumber  \\
&&\times  \int dE_e P_{\nu,{\rm syn}}(\nu,E_{e})E_e^2 \left[\frac{f_{e^{\pm}}(E_e-h\nu)}{(E_e-h\nu)^2} - \frac{f_{\rm e^{\pm}}(E_e)}{E_e^2}\right]\ , \nonumber \\
\label{eq:chi_syn}
\end{eqnarray}
where $F(\nu/\nu_c)$ is the synchrotron  spectrum for an electron (positron) with  characteristic frequency $\nu_c = 3\gamma_e^2 e B /(4\pi m_e c)$, and $P_{\nu, {\rm syn}}$ is its specific power; $f_{e^{\pm}}$ is the energy distribution of electrons (positrons), which effectively depends on the plasma temperature ($\hat{T}$) and is assumed to follow the Maxwell–J$\ddot{\text{u}}$ttner distribution~\cite{1911AnP...339..856J}; $E_e = \gamma_e m_e c^2$ represents the  electron (positron) energy. 
The total emission and absorption coefficients are given by the sum of those for each radiation mechanism.

\subsection{\label{sec:subshock} Subshock formation} 
The results presented in Ref.~\cite{2010ApJ...725...63B}, using  the governing equations introduced in Sec.~\ref{sec:dynamics} for $\sigma_u = 0$, show that the discontinuity of the the bulk motion velocity at the shock front is negligible, with particle acceleration being inefficient. We contrast this finding with the case of a magnetized fluid--see also Refs.~\cite{2017ApJ...838..125B,2019ApJ...879...83L}. In the latter scenario, the compressed magnetic field stores a fraction of the total energy and pressure, which is not converted into radiation, and a subshock embedded in the radiation-mediated  shock forms. 

In order to obtain a subshock where particles can be efficiently accelerated, the magnetic energy density in the shock downstream should dominate over the radiation energy density. 
However, if the magnetization is too high, a shock discontinuity cannot form efficiently. In the far upstream, the condition $\Gamma_u > \Gamma_{\rm F,max}$ (see Eq.~\ref{eq:FMS}) defines a critical  $\sigma_c = \Gamma_u^2 - 1$. For $\sigma_u \sim \sigma_c$,  a weak shock develops with  negligible velocity jump, whereas for $\sigma_u > \sigma_c$, a rarefaction wave forms instead of a shock.

In the rest of this section, we neglect the radiation feedback for simplicity. The thermal energy represents the total energy budget that could go into radiation. The specific enthalpy of the magnetized fluid is
\begin{eqnarray}
    h = 1 + w + \sigma\, ,
\end{eqnarray}
where 
$w = {(e_{\rm fl} + P_{\rm fl})}/{(n_p m_p c^2)}$
accounts for thermal energy and pressure. In the downstream, the ratio between the enthalpy due to  thermal energy and pressure and the magnetic field is 
\begin{eqnarray}
    \frac{w_d}{\sigma_d} = \left[1+\frac{3}{2}f(T_d)\right]\frac{m_e}{m_p} \hat{T}_d \frac{1}{\sigma_d}\, .
\end{eqnarray}

\begin{figure}
\centering
\includegraphics[width=0.99\columnwidth]{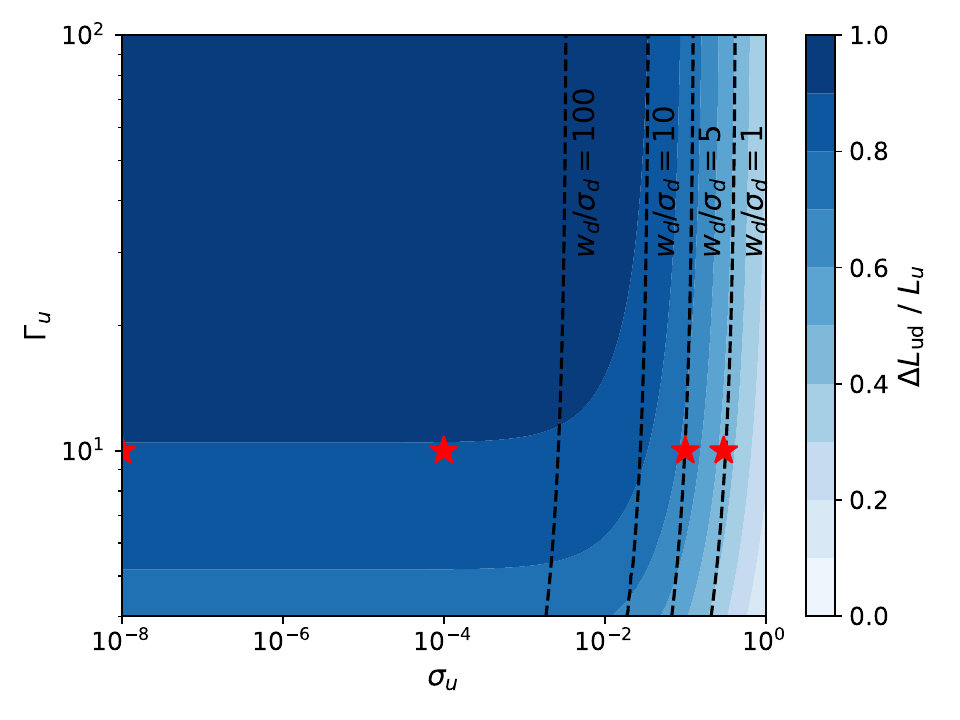}
\caption{Isocontours of 
$w_d/\sigma_d$ (dashed black lines) in the plane spanned by  $\sigma_u$ and $\Gamma_u$. The color scale represents the contours of the energy budget converted from kinetic energy into other forms of energy ($L_{\rm TRP}/L_u$).  The red stars mark the representative configurations adopted in this work. For fixed $\Gamma_u$, $w_d/\sigma_d$ and $\Delta L_{\rm ud}/L_u$ decrease as $\sigma_u$ increases.}
\label{fig:sim_para}.
\end{figure}

When crossing the shock, the amount of  kinetic energy of the upstream converted into thermal energy (radiation) that could be used to accelerate particles is
\begin{eqnarray}
\Delta L_{\rm ud} &=& L_u - L_d \nonumber \\
&=& \left[\frac{1}{1+\sigma_u}\left(1-\frac{\Gamma_d}{\Gamma_u}\right) + \left(1-\frac{\beta_u}{\beta_d}\right)\frac{\sigma_u}{1+\sigma_u}\right]L_u\, . \nonumber \\
\label{eq:total_energy_budget}
\end{eqnarray}
Here, $L_u = \Gamma_u^2 \beta_u \rho_u c^2(1+\sigma_u)$ and $L_d = \Gamma_d^2 \beta_d \rho_d c^2 (1+\sigma_d)$ are the upstream and downstream kinetic energy fluxes, respectively. To determine the region of the parameter space where a subshock can exist, we solve  Eqs.~\ref{eq:B_flux_cons}, \ref{eq:E_cons_normal}, and \ref{eq:M_cons_normal}, without radiation feedback. This approach allows us to compute the bulk motion velocity, temperature, and  magnetization parameter in the downstream. 

Figure~\ref{fig:sim_para} shows isocontours of $w_d/\sigma_d$ and $\Delta L_{\rm ud}/L_u$ in the parameter space spanned by $\Gamma_u$ and $\sigma_u$. To retain a subshock in the presence of a radiation-mediated shock, the condition $w_d/\sigma_d \lesssim 1$ must be satisfied (this implies that the magnetic energy density dominates over or is comparable to the thermal energy). We can see from Fig.~\ref{fig:sim_para} that $w_d/\sigma_d \gtrsim 1$ for a large fraction of our parameter space.

In the following, we explore the evolution of the shock structure for five benchmark configurations in the parameter space shown in Fig.~\ref{fig:sim_para}, all with  $\Gamma_u = 10$. We select $\sigma_u = 0$, $10^{-8}$, $10^{-4}$, $\sigma_u = 0.1$ and $0.3$ to explore the transition from non-magnetized to mildly magnetized regimes. Such configurations are marked with  red stars in Fig.~\ref{fig:sim_para}; we can see that  both $w_d/\sigma_d$ and $\Delta L_{\rm ud}/L_u$ decrease as $\sigma_u$ increases.

\section{Numerical setup}
\label{sec:numerics}

In this section, we present the iterative approach adopted to solve the hydrodynamic and kinetic equations  for the steady-state shock solution numerically. We consider two angular grids, one for the upstream-going directions and one for the downstream-going directions, following Ref.~\cite{2010ApJ...725...63B}. For the upstream-going directions ($0 \leq \mu_{\rm sh} \leq 1$), we assume that $\mu_{\rm sh}$ is binned logarithmically; for the downstream-going directions ($-1 \leq \mu_{\rm sh} < 0$), we use the Gaussian quadrature (cf.~Appendix D in Ref.~\cite{2010ApJ...725...63B} for details). The number of angular bins is $N_{\mu_{\rm sh}} = 40$. The photon frequency grid is chosen to be logarithmic, ranging from $\hat{\nu}_{\rm sh} = 10^{-9}$ to $\hat{\nu}_{\rm sh} = 10^{14}$ with  $N_{\nu_{\rm sh}} = 200$ bins. 

The position of each fluid element is denoted by the normalized optical depth, with $\hat{\tau} = 0$ at the shock. In the downstream region, we adopt $d\hat{\tau} = 0.1$ for all our benchmark configurations except for $\sigma_u =0.3$, for which we use $d\hat{\tau}=0.0025$. In the far downstream, in principle, one should apply a reflective boundary, such that  the upstream- and downstream-going photon intensities in the comoving frame of the fluid are identical. This condition reflects the fact that  photons and leptons are expected to reach thermal equilibrium. However, for no ($\sigma_u = 0$) and small magnetization ($\sigma_u = 10^{-8}$ and $10^{-4}$), the downstream-going photon frequencies are boosted by the relativistic bulk motion of the upstream flow; hence, photons cannot be fully absorbed over short distances. 
Therefore, we adopt an open boundary at $\hat{\tau} = 50$. For the configurations with $\sigma_u = 0.1$ and $0.3$, we use a reflective boundary at $\hat{\tau} = 5$. This is because the  synchrotron cooling timescale of pairs in the highly magnetized case is much shorter than that of other processes, and the plasma and radiation reach equilibrium more quickly in the downstream. High spatial resolution is required in these cases, and we find that a reflective boundary optmizes  the simulation convergence time. In the upstream region, $d\hat{\tau}$ is uniformly binned in logarithmic space, with $N_{\hat{\tau}} = 500$ over $-500 < \hat{\tau} < 0$; but for  $\sigma_u = 0.1$ and $0.3$, we use $N_{\hat{\tau}} = 100$ over $-100 < \hat{\tau} < 0$. As for the far upstream, our numerical solution can extend up to large distances from the shock discontinuity, where radiation has been fully absorbed. Hence, we adopt $\hat{I}_{\hat{\nu}_{\rm sh}}(\mu_{\rm sh}) = 0$ and $a_{e^{\pm}} = 0$ at the upstream  boundary  (cf.~Fig.~\ref{fig:sketch}).

\begin{figure*}
\centering
\includegraphics[width = 1.98\columnwidth]{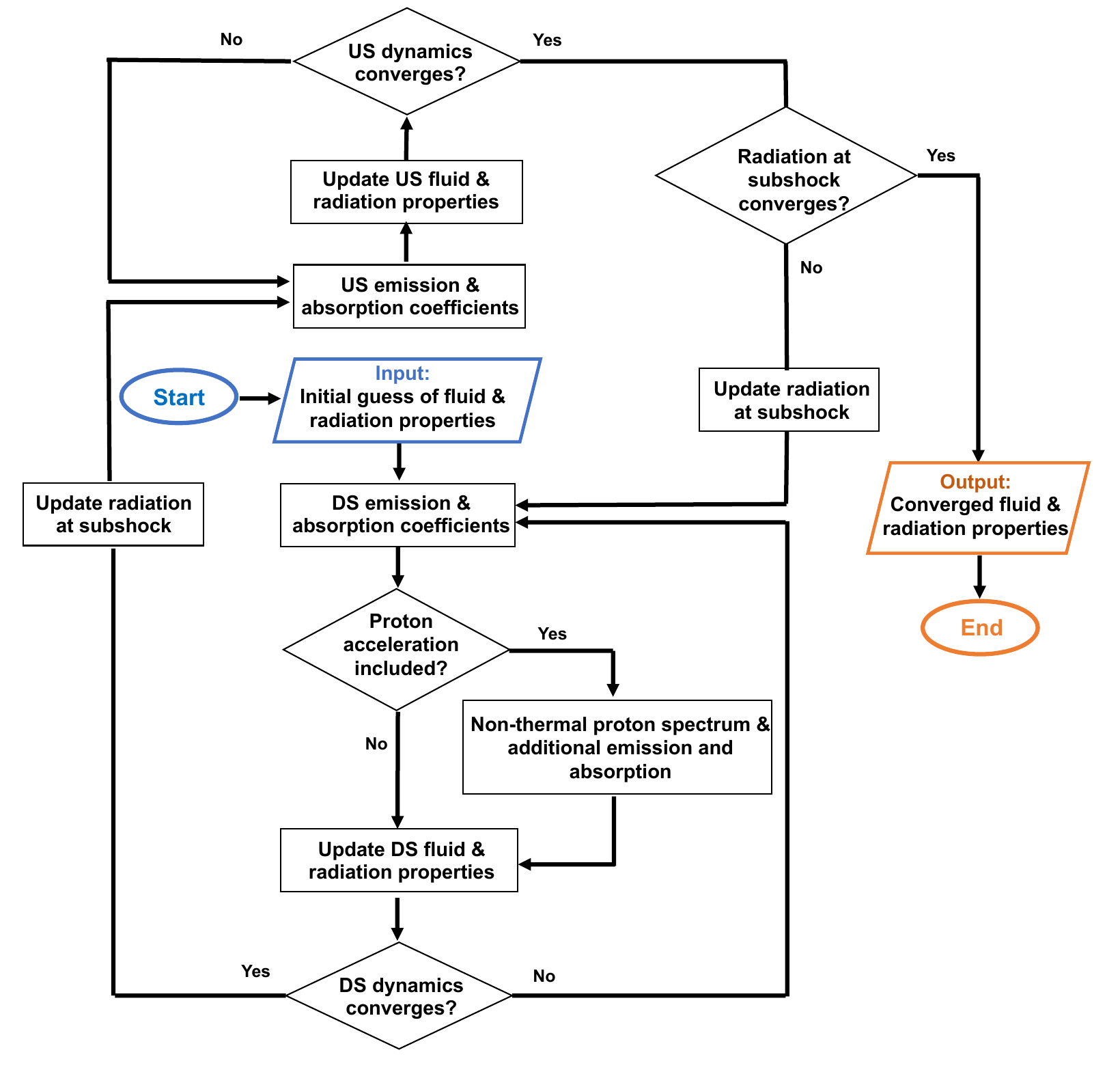} 
\caption{Flow chart summarizing the iterative approach adopted to solve the hydrodynamic and kinetic equations  for a steady-state shock. We follow the evolution of the upstream (US) and downstream (DS) fluid and radiation properties, the related shock structure, as well as the photon spectral distribution accounting for leptonic and hadronic processes.}
\label{fig:sim_procedure}
\end{figure*}

For each of our benchmark $(\sigma_u, \Gamma_u)$, the simulation starts from the downstream. First, we make an initial guess for $I(\mu_{\rm sh},\nu_{\rm sh}, \hat{\tau})$ and $a_{e^{\pm}} (\hat{\tau})$. We then solve Eqs.~\ref{eq:E_cons_normal} and \ref{eq:M_cons_normal} and select the subsonic solution. This procedure allows to obtain the temperature ($\hat{T}$) and Lorentz factor ($\Gamma$) profiles. Then the emission and absorption coefficients (Eqs.~\ref{eq:eta_s}--\ref{eq:chi_pair} and \ref{eq:eta_ff}--\ref{eq:chi_syn}) are calculated and the radiation transfer equation (Eq.~\ref{eq:radiation_transfer_normal}) is solved. Next, we update the intensity and $a_{e^{\pm}}$ at each location $z (\hat{\tau})$, cf.~Fig.~\ref{fig:sketch}, relying on the characteristic quantities computed in adjacent bins, according to Eqs.~\ref{eq:radiation_transfer_normal} and \ref{eq:pair_prodution_normal}. This procedure is repeated with the updated $I(\mu_{\rm sh},\nu_{\rm sh}, \hat{\tau})$ and $a_{e^{\pm}} (\hat{\tau})$ profiles until convergence is achieved. We then use the converged $\hat{I}(\mu_{\rm sh} < 0)$ at $\hat{\tau} = 0$ as the boundary condition for the upstream  and  iterate using the same procedure described above while selecting the supersonic solution until convergence is reached. The converged $\hat{I}(\mu_{\rm sh} > 0)$ at $\hat{\tau} = 0$ is then adopted as the boundary condition for the downstream. The upstream and downstream simulations are repeated alternately until the entire system reaches convergence. A flow chart summarizing our numerical approach is shown in Fig.~\ref{fig:sim_procedure}.

\section{Magnetized shocks mediated by radiation produced by leptonic processes}
\label{sec:RMSleptonic}
\begin{figure*}
\centering
\includegraphics[width = 1.98\columnwidth]{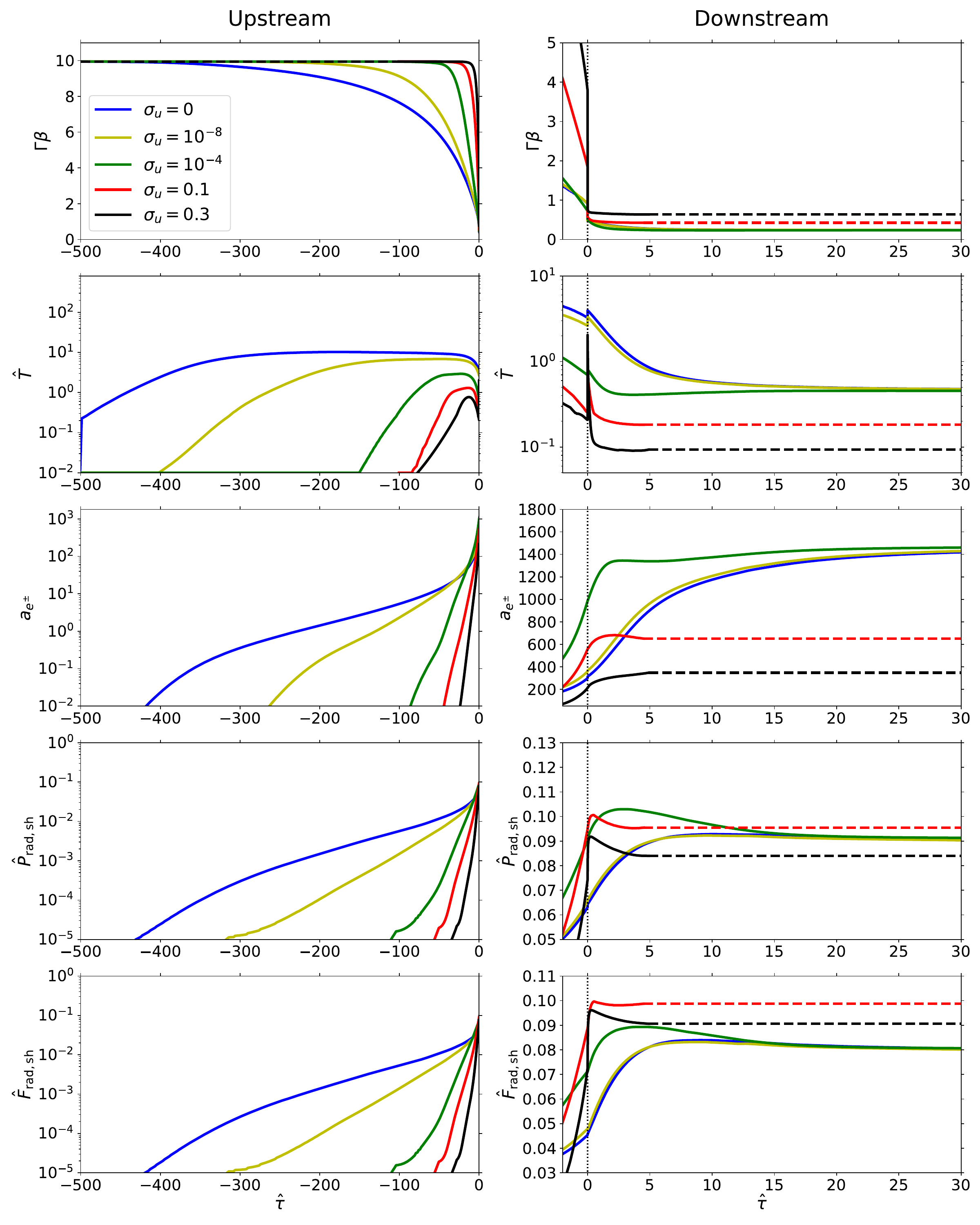}
\caption{Bulk Lorentz factor, plasma temperature, fraction of positrons, radiation pressure, and radiation flux as functions of $\hat{\tau}$, from top to bottom and for our five benchmark configurations (cf.~Fig.~\ref{fig:sim_para}): $\sigma_u = 0$ in blue, $\sigma_u = 10^{-8}$ in yellow, $\sigma_u = 10^{-4}$ in olive green, $\sigma_u = 0.1$ in red, and $\sigma_u = 0.3$ in black, respectively. The upstream (downstream) profiles are shown in the left (right) panels. As evident from the right panels, for $\sigma_u \neq 0$, a subshock forms and becomes more prominent as $\sigma_u$ increases; its position is marked with a vertical dotted line.}
\label{fig:lep_profile}.
\end{figure*}

First, we investigate the impact of radiation on the shock structure by considering leptonic processes only.
Figure~\ref{fig:lep_profile} displays the profiles of the bulk Lorentz factor ($\Gamma$), plasma temperature ($\hat{T}$), fraction of  positrons ($a_{e^{\pm}}$), radiation pressure, and radiation flux for our benchmark simulations with $\sigma_u = 0$, $10^{-8}$, $10^{-4}$, $0.1$, and $0.3$. For $\sigma_u = 0$, the profile of the bulk Lorentz factor $\Gamma$ is identical to the one of  Ref.~\cite{2010ApJ...725...63B}. However, our simulation leads to a lower pair fraction $a_{e^{\pm}}$ and a higher plasma temperature $\hat{T}$ at the subshock by a factor of $\sim 3$, with similar equilibrium values in the far downstream region as those in Figs.~9 and 11 of Ref.~\cite{2010ApJ...725...63B} ($\hat{T}\simeq 200~{\rm keV}$ and $q_{e^\pm} \simeq 10^3$). These discrepancies between our findings and the ones of Ref.~\cite{2010ApJ...725...63B}  are due to differences in the pair production rate, but   the production of $a_{e^{\pm}}$ and $\hat{T}$, which reflects the plasma thermal pressure, is consistent between our results and theirs.
Comparable  equilibrium values are found in Ref.~\cite{2018MNRAS.474.2828I} using a Monte Carlo approach.

The subshock jump is negligible for $\sigma_u = 0$, $10^{-8}$, and $10^{-4}$. However, it tends to increase with increasing $\sigma_u$.
For $\sigma_u > 0.1$, a prominent discontinuity in $\Gamma$ and $\hat{T}$ occurs at $\hat{\tau} = 0$, whereas the profiles of $a_{e^{\pm}}$, $\hat{P}_{\rm rad,sh}$, and $\hat{F}_{\rm rad,sh}$ remain continuous.  These results confirm the early findings of  Ref.~\cite{2017ApJ...838..125B}. Although the  magnetization does not significantly affect the downstream fluid dynamics, especially in the weakly magnetized cases, it  has a substantial impact on the upstream shock structure. 
In particular, although synchrotron radiation is subdominant compared to other radiative processes, synchrotron self-absorption  plays a significant role for  the benchmark configurations with $\sigma_u \neq 0$ (cf.~Appendix~\ref{sec:lep_contribution}). Initially, the upstream fluid is cool, allowing for the upstream-going photons to be Compton scattered into a relatively low-energy range and  absorbed by leptons via synchrotron self-absorption. The heated leptons transfer energy to protons via Coulomb scattering, decelerating the inflowing upstream fluid. In the absence of  magnetization, the photon–lepton energy transfer is less efficient, hence the upstream-going photons propagate efficiently and the upstream profile becomes less sharp. We refer the interested reader to Appendix~\ref{sec:lep_contribution} for more details.

\begin{figure*}
\centering
\includegraphics[width = 1.98\columnwidth]{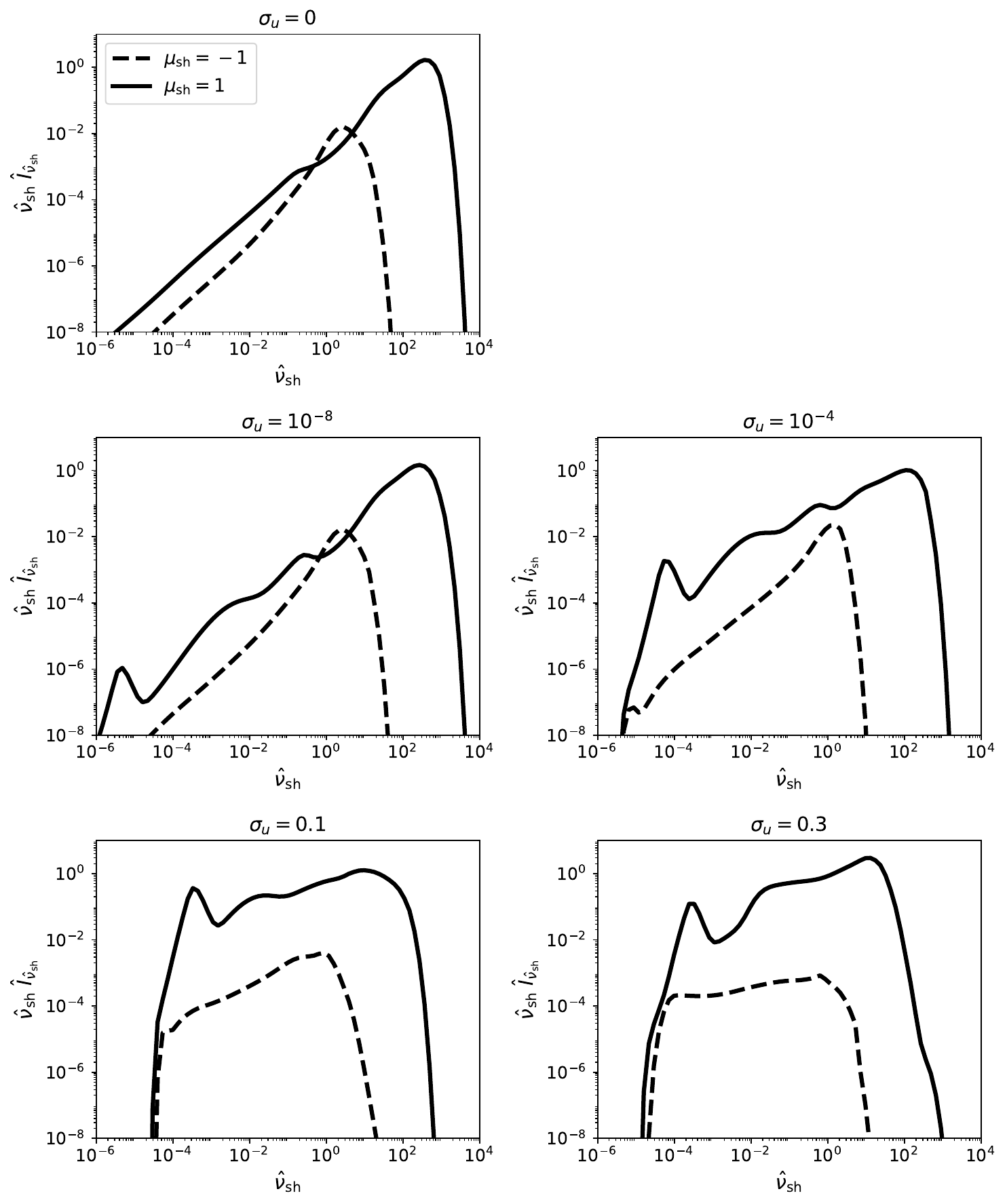}
\caption{Photon spectral intensity at the subshock for our five benchmark configurations with $\Gamma_u = 10$ and $\sigma_u = 0$, $10^{-8}$, $10^{-4}$, $0.1$, and $0.3$ from top left to bottom right, respectively. The radiation intensities for the  upstream-going ($\mu_{\rm sh} = -1$) and downstream-going ($\mu_{\rm sh} = 1$) directions are plotted with dashed and solid lines, respectively. For $\sigma_u \lesssim 10^{-4}$, the peak frequency of the downstream photon spectrum is about two orders of magnitude larger than the one of the upstream spectrum because the upstream temperature and bulk Lorenz factor are larger than the downstream ones. For $\sigma_u \gtrsim 0.1$, the photon spectrum flattens due to non-negligible synchrotron losses. }
\label{fig:lep_spectrum}
\end{figure*}

Once a subshock forms, the jump in the bulk flow velocity between the upstream and downstream regions can facilitate particle acceleration. The resulting photon spectrum at the subshock is therefore modified, as illustrated in Fig.~\ref{fig:lep_spectrum}, where the radiation intensity is plotted for the upstream- ($\mu_{\rm sh} = -1$) and downstream-going ($\mu_{\rm sh} = 1$) directions. For $\sigma_u \lesssim 10^{-4}$, the peak frequency of downstream-going photons is   larger by a factor $\Gamma_u^2$ (i.e., about two orders of magnitude) than the one of the upstream-going ones (cf.~solid vs.~dashed curves in Fig.~\ref{fig:lep_spectrum}). This is because: (1) The upstream temperature ($\hat{T} \simeq 10$) is larger than the downstream one ($\hat{T} \lesssim 1$); hence, photons newly generated or scattered in the upstream, and contributing to the downstream-going spectrum at the subshock, tend to have higher frequencies. (2) The fluid in the upstream has a higher bulk Lorentz factor, which boosts the photon frequency to a higher value. For $\sigma_u \gtrsim 0.1$, the spectrum of upstream-going photons tends to become flatter at $\hat{\nu}_{\rm sh} \lesssim 1$, and the spectrum of downstream-going photons  flattens at $\hat{\nu}_{\rm sh} \lesssim 10^2$. This happens because synchrotron radiation in the downstream region becomes non-negligible in this energy range, and the interplay between synchrotron self-absorption and Compton scattering in the upstream populates this energy range with photons. The downstream-going photons further affect the upstream-going photons, enhancing the spectral  flattening. The cutoff at $\hat{\nu}_{\rm sh} \simeq 10^{-4}$ is due to strong synchrotron self-absorption at low frequency.

We assume $n_{p,u} = 10^{15}~{\rm cm}^{-3}$ for the upstream proton number density. Although all physical quantities relevant for the shock are normalized so that they are independent of $n_{p,u}$, the synchrotron radiation and self-absorption depend on $n_{p,u}$. For smaller $n_{p,u}$, we expect that the synchrotron characteristic frequency should become smaller, with  consequent less significant impact of the magnetization  on the shock dynamics. In fact, since $\nu_c \propto B \propto (\sigma n_p)^{1/2}$, we expect that the magnetic field should have a significant influence on the shock dynamics for $\sigma_u (n_{p,u}/10^{15}{\rm cm}^{-3})^{-1} \gtrsim 10^{-8}$.

\section{\label{sec:RMShadronic}Magnetized shocks mediated by radiation produced by  leptonic and hadronic processes}
The inelastic interactions between accelerated and background protons ($pp$ interactions), as well as those between accelerated protons and photons ($p \gamma$  interactions), can alter the photon spectrum, thereby influencing the radiation flux and pressure, and ultimately modifying the profiles of the physical quantities characterizing the shock. In this section, we outline our treatment for $pp$ and $p\gamma$ interactions, introducing the corresponding acceleration and cooling rates. We then investigate the impact of hadronic processes on the shock structure and photon distribution.

\subsection{\label{sec:particle_acceleration}Proton acceleration}
Proton acceleration is expected to be more efficient than lepton acceleration~\cite{2011ApJ...726...75S}. The proton acceleration rate is 
\begin{eqnarray}
t_{\rm acc}^{-1} = \frac{qB}{\gamma_p m_p c}\, ,
\label{eq:p_acc}
\end{eqnarray}
where $q = 4.8 \times 10^{-10} {\rm esu}$ is the elementary charge and $\gamma_p$ is the proton Lorentz factor. 
All quantities are defined in the comoving frame of the immediate downstream fluid.

Accelerated protons cool through several mechanisms. Here, we consider inverse Compton scattering, synchrotron radiation, inelastic  $pp$ collisions, and $p\gamma$ interactions. The maximum energy up to which protons can be accelerated ($E_{p,\rm{max}}$) is obtained by equating the acceleration rate to the cooling rate: $t_{\rm acc}^{-1} = t_{\rm cool}^{-1}$. The latter is given by the sum of the cooling rates characteristic of the different radiative processes, as outlined in  Appendix~\ref{app:acc_cool}. Since the acceleration time is assumed to be in the Bohm regime (i.e., we work in the most efficient scenario), our maximum proton energy represents an upper limit. However, for a power-law energy distribution with index $p > 2$, $E_{\rm p,max}$ does not affect the total energy or number of accelerated protons. A lower $E_{\rm p,max}$ would therefore only reduce the ultra-high-energy radiation without altering our conclusions on the structure of the shock.

Assuming that a fraction ($\epsilon_p = 0.1$~\cite{2009ApJ...698.1523S,2011ApJ...726...75S}) of kinetic energy of the bulk fluid motion is converted into proton acceleration, and taking into account the kinetic energy loss at the subshock (analogous to Eq.~ \ref{eq:total_energy_budget}), the resultant luminosity of protons  undergoing acceleration is
\begin{widetext}
\begin{eqnarray}
L_{\rm p,acc} = \epsilon_p (L_{\rm bulk,\hat{\tau}_0^-} - L_{\rm bulk,\hat{\tau}_0^+}) = \epsilon_p L_{u,p} \left\{ \frac{\Gamma_{\hat{\tau}_0^-} - \Gamma_{\hat{\tau}_0^+}}{\Gamma_u}\left[1 + \frac{m_e}{m_p}(2a_{e^{\pm},\hat{\tau}_0}+1)\right] + \sigma_u \beta_u \left(\frac{1}{\beta_{\hat{\tau}_0^-}} - \frac{1}{\beta_{\hat{\tau}_0^+}}\right) \right\}\, ,
\label{eq:L_acc}
\end{eqnarray}
\end{widetext}
where the luminosity of the bulk flow is
\begin{eqnarray}
L_{\rm bulk} &=& \Gamma^2 \beta c^3 A n_p m_p \left[1 + \frac{m_e}{m_p}(2a_{e^{\pm}} + 1)\right]  \\
&&+ \Gamma^2 \beta c^3 A n_p m_p \sigma\, ; \nonumber
\end{eqnarray}
the subscripts ``$\hat{\tau}_0^{+}$'' and ``$\hat{\tau}_0^{-}$'' denote the physical quantities immediately downstream and upstream of the subshock, respectively, and $A$ is the arbitrary area  assumed for the shock plane. The luminosity of the bulk proton flow in the far upstream is
\begin{eqnarray}
L_{p,u} = \Gamma_u^2 \beta_u n_{p,u} m_p c^3 A \, .
\label{eq:Lu}
\end{eqnarray}
Note that the value of $A$ does not affect our results, since only the proton number density enters our simulations.

We assume that the energy distribution of accelerated protons follows a broken power law:
\begin{eqnarray}
n_{p,{\rm acc}}(E_p) \propto E_p^{-2} \exp\left(-\frac{E_p}{E_{\rm p,max}}\right)\, ,
\label{eq:nonthermal_p_spec}
\end{eqnarray}
normalized such that
\begin{eqnarray}
\Gamma_{\hat{\tau}_{0}^{+}} \beta_{\hat{\tau}_0^{+}} cA \int_{E_{\rm p,min}}^{+\infty} n_{p,{\rm acc}}(E_p) E_p dE_p = L_{\rm p,acc}\ .
\end{eqnarray}
Moreover,  we consider $E_{\rm p, min} = m_p c^2$. 

\subsection{\label{sec:had_feedback} Method to account for hadronic processes}

To assess the impact of $pp$ and $p \gamma$ interactions, we solve Eqs.~\ref{eq:B_flux_cons} and \ref{eq:M_cons_normal}--\ref{eq:pair_prodution_normal}, while modifying the emission and absorption coefficients in the radiation transfer equation (Eq.~\ref{eq:radiation_transfer_normal}) to account for these additional interaction channels. We assume: (1) accelerated protons move with the  bulk velocity of the fluid, but undergo random motions in the fluid comoving frame; (2) hadronic collisions are isotropic in the fluid comoving frame; and (3) accelerated protons are confined to the downstream region. 
Moreover, the non-thermal protons are not thermalized by interactions with the medium  and, for $\epsilon_p = 0.1$, only a small fraction of protons is accelerated.

For the downstream, we  compute the spectrum of non-thermal photons at each position. Then, we obtain the emission and absorption coefficients for $pp$ and $p \gamma$  interactions employing the open-source software $\text{AM}^3$~\cite{2024ApJS..275....4K} (whereas the emission and absorption coefficients for the leptonic processes are computed as illustrated in Sec.~\ref{sec:radiation}).
To this purpose, the non-thermal proton spectrum at the subshock (Eq.~\ref{eq:nonthermal_p_spec}) is used as an input for the AM$^3$ simulation, together with the photon spectrum at the subshock obtained considering leptonic processes. We account for  radiative processes linked to accelerated protons (i.e., Compton scattering, synchrotron), as well as pion and muon decays. The number density of background protons and the strength of the background magnetic field are set according to Eqs.~\ref{eq:p_number_cons} and \ref{eq:B_flux_cons}, respectively. Through this approach, we follow the evolution of the non-thermal proton spectrum from the subshock to the downstream boundary~\footnote{Note that this effectively corresponds to follow the evolution of protons in a Lagrangian frame comoving with the fluid; as a consequence,  volume expansion and particle escape are not accounted for.}. The overall simulation time is determined by the size of the shock region,  with the downstream region extending up to $\hat{\tau}_{\rm end}$:
\begin{eqnarray}
T_{\rm sim, tot} = \int_0^{\hat{\tau}_{\rm end}} \frac{d\hat{\tau}}{\Gamma^2 (1+\beta)n_p(1 + 2{a_{e^{\pm}}})\sigma_T \beta c}\ .
\label{eq:T_sim_tot}
\end{eqnarray}
The non-thermal proton spectrum at any position ($\hat{\tau}$) is computed at intermediate simulation time. 

At the far-downstream boundary (see Fig.~\ref{fig:sketch}),  reached by protons within $T_{\rm sim, tot}$, non-thermal protons may not have fully cooled. However, the photons emitted by  accelerated protons in the far downstream may propagate back and affect  the photon spectrum as well as the fluid dynamics near the subshock. To account for this contribution,  we introduce a ghost cell beyond the downstream boundary (whose  size,  $\hat{\tau}_{\rm gh, end}$, is larger than the characteristic scale of all relevant radiative processes). The escape time (i.e., the time the fluid takes to travel through the ghost cell)  is calculated using an equation similar to Eq.~\ref{eq:T_sim_tot}, with $\hat{\tau}_{\rm end}$ being replaced by $\hat{\tau}_{\rm gh, end}$. We inject protons according to the spectrum at the downstream boundary (obtained in the previous step) and we inject photons according to the spectrum at the downstream boundary computed using leptonic processes only. Their injection rates are determined by the speed of light and the bulk velocity, respectively. We then re-run  the AM$^3$ simulations in the ghost cell until the output photon spectrum converges. The resulting photon spectrum is used to compute the intensity in the ghost cell, whose angular dependence in the shock frame is determined by the bulk fluid motion. This intensity, together with that of the reflected leptonic radiation, is injected into the main simulation region from the downstream boundary.

After obtaining the non-thermal proton spectrum at each position and initializing the ghost cell as illustrated above, we account for $pp$ and $p\gamma$ interactions in the downstream, following Fig.~\ref{fig:sim_procedure}. We compute the emission and absorption coefficients running  the AM$^3$ code for each  $\hat\tau$ in the downstream.
Photons are  injected in AM$^3$  with a number density and energy distribution derived from the local radiation intensity, 
\begin{eqnarray}
n_{\rm ph,\nu} = \frac{2\pi}{hc\nu}\int I_{\hat{\nu}}(\mu) d\mu\, ,
\end{eqnarray}
where $n_{\rm ph,\nu}$ is the photon number density, and $I_{\hat{\nu}}(\mu)$ is the photon intensity, updated at each iteration step. For simplicity, we compute the emission and absorption coefficients  for each  process separately, each with an arbitrary simulation time $T_{\rm sim}$. The emission and absorption coefficients for hadronic processes (with ``$i$'' standing for ``$pp$,'' ``$p\gamma$,'' and ``syn'') are 
\begin{eqnarray}
&&\hat{\eta}_{i,\rm{sh}}(\hat{\nu}_{\rm sh},\mu_{\rm sh}) = \left[\Gamma(1-\beta\mu_{\rm sh})\right]^{-2}  \\
&&~~~~~~\times \frac{1}{\Gamma (1+\beta)\sigma_T (1 + 2a_{e^{\pm}})n_p} \frac{m_p}{m_e\Gamma_u^2 \beta_u n_{p,u} hc} \frac{I_{e,\hat{\nu},i}}{c T_{\rm sim}}\, , \nonumber 
\end{eqnarray}
\begin{eqnarray}
&&\hat{\chi}_{i,{\rm sh}}(\hat{\nu}_{\rm sh},\mu_{\rm sh}) = \Gamma(1-\beta \mu_{\rm sh})  \\
&&~~\times \frac{1}{\Gamma(1+\beta)\sigma_T(1 + 2a_{e^{\pm}})n_p} \frac{I_{\rm in,\hat{\nu},i} - (I_{\rm out,\hat{\nu},i} - I_{e,\hat{\nu},i})}{cT_{\rm sim}}\, , \nonumber 
\end{eqnarray}
where $I_{{\rm in},\hat{\nu},i}$ represents the intensity of the input radiation;  $I_{{\rm out},\hat{\nu},i}$ and $I_{e,\hat{\nu},i}$ are the total radiation intensity and the newly emitted radiation intensity, respectively, at the end of the simulation. Moreover, $\hat{\chi}_{pp, {\rm sh}} = \hat{\chi}_{\rm syn, sh} = 0$ at all angles and frequencies. Note that $I_{{\rm in},\hat{\nu},i}$, $I_{{\rm out},\hat{\nu},i}$, and $I_{e,\hat{\nu},i}$ are assumed to be isotropic (for consistency with  the isotropy assumption on the emission and absorption coefficients in the fluid comoving frame). The
photon number density used as  input/output of the AM$^3$ code is then computed  using the fact that $I_{\hat{\nu}} = {c n_{\rm ph,\nu} h\nu}/{(4\pi)}$.
Finally, the interplay between  hadronic and leptonic processes is accounted for by Eq.~\ref{eq:radiation_transfer_normal}.

In summary, the module AM$^3$  interfaces our hydrodynamical and radiation transfer modules multiple times for different purposes in the downstream. We first  obtain the non-thermal proton spectrum for all  $\hat{\tau}$'s. Then, we compute  the cooling of non-thermal protons, in order to determine the net inflow of photons from the downstream boundary. After these steps, we call the module AM$^3$ at each downstream position to compute the emission and absorption coefficients. At the same time, the shock solution  for the downstream region is computed. After the downstream solution converges, we use the photon spectrum at the subshock as boundary condition for the upstream. We then perform iterations in the upstream region in the presence of $p\gamma$ interactions and leptonic processes.

\subsection{Impact of  hadronic processes on the shock structure and photon spectrum}
\label{sec:results_had}
We investigate the shock dynamics in the presence of hadronic and leptonic processes for our benchmark configuration with $\sigma_u = 0.1$. We select this representative $\sigma_u$, since a prominent subshock appears (cf.~Fig.~\ref{fig:lep_profile}), and the magnetization is not so high as to hinder the efficiency of particle acceleration \cite{2017ApJ...838..125B}. 

\begin{figure*}
\centering
\includegraphics[width = 1.98\columnwidth]{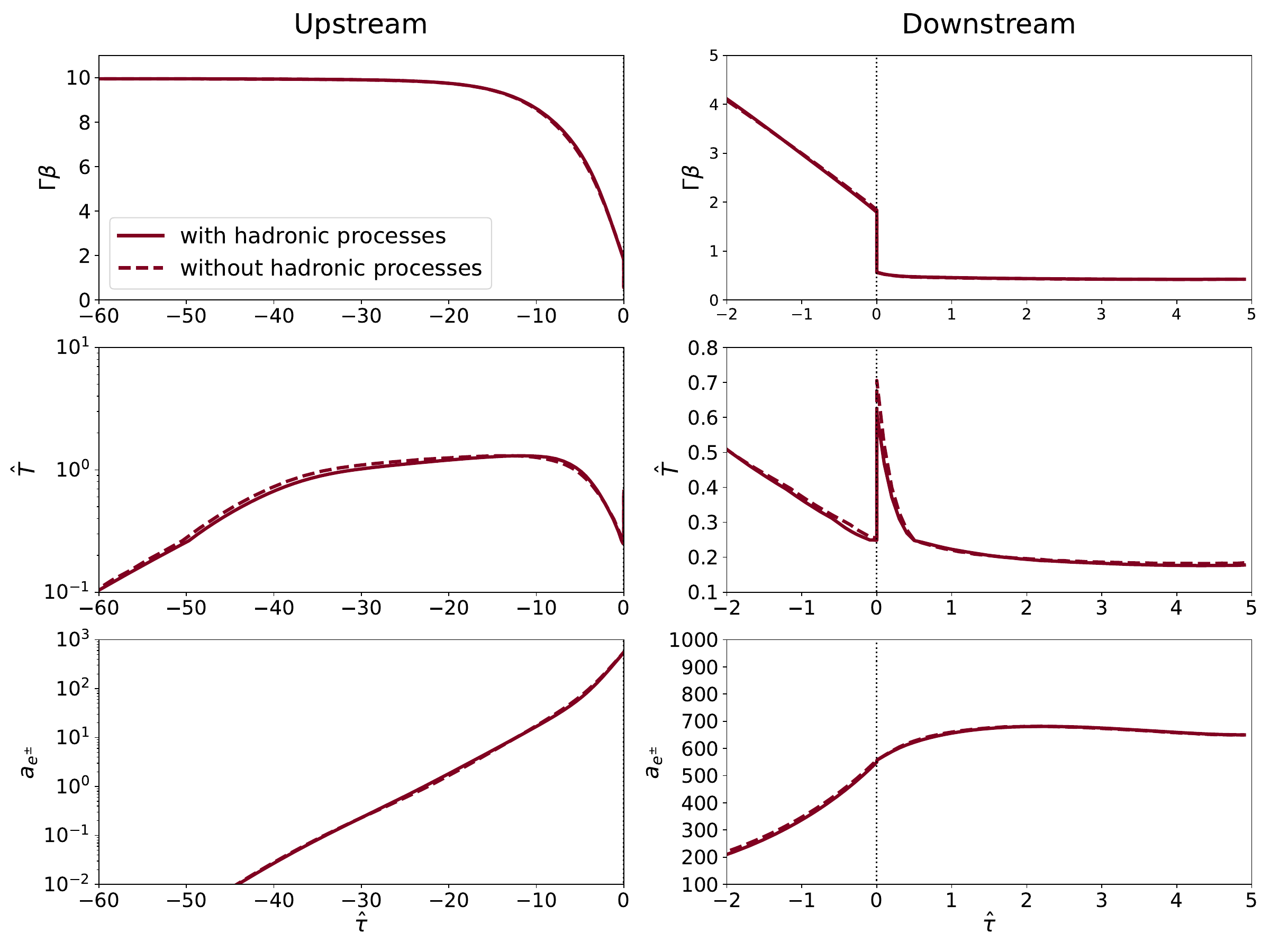}
\caption{Same as the top three panels of Fig. \ref{fig:lep_profile}, but  for $\sigma_u = 0.1$. The solid (dashed) lines represents the bulk Lorentz factor, the plasma temperature and the positron fraction with  (without) $pp$ and $p\gamma$ interactions, respectively. Hadronic processes are responsible for  minimal variations with respect to the scenario when only leptonic processes are taken into account.}
\label{fig:had_profile}
\end{figure*}

Figure~\ref{fig:had_profile} displays the profiles of the bulk Lorentz factor, plasma temperature, and fraction of positrons, with and without hadronic processes. We can see that $pp$ and $p \gamma$ interactions negligibly contribute to the total radiation flux and pressure. Hence,  the shock structure is mostly impacted by leptonic interactions. This is also due to the fact that the inelastic collisions between background protons and photons produced by leptonic processes in the upstream are negligible, because their interaction length scale is much larger than that of  leptonic processes ($\hat{\tau}_{p\gamma} \simeq \Gamma(1+\beta)(1+2a_{e^{\pm}})(n_p/n_u)(\sigma_{p\gamma}/\sigma_T) \gtrsim  10^3$, where $\sigma_{p\gamma} = 5\times 10^{-28}~{\rm cm^2}$ represents the peak $p\gamma$ cross section). 

\begin{figure}
\centering
\includegraphics[width = 0.99\columnwidth]{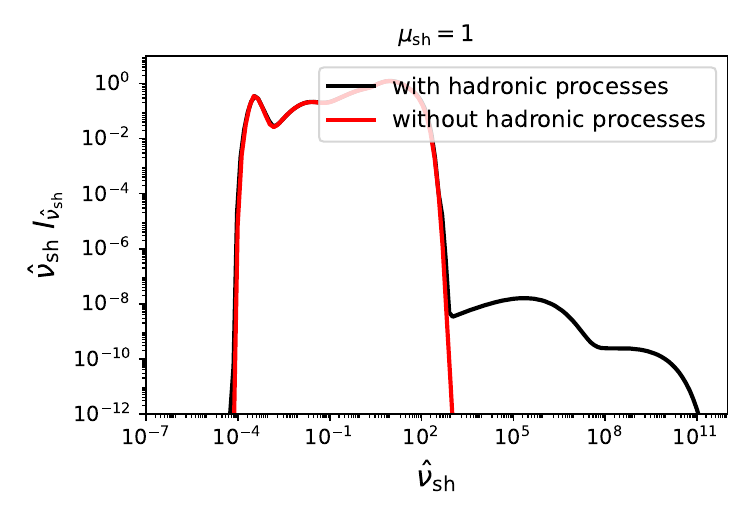} \\
\includegraphics[width = 0.99\columnwidth]{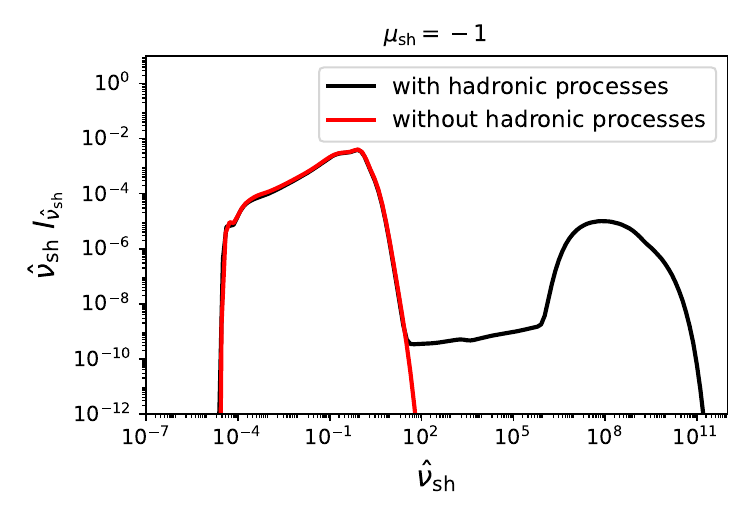} \\
\includegraphics[width = 0.99\columnwidth]{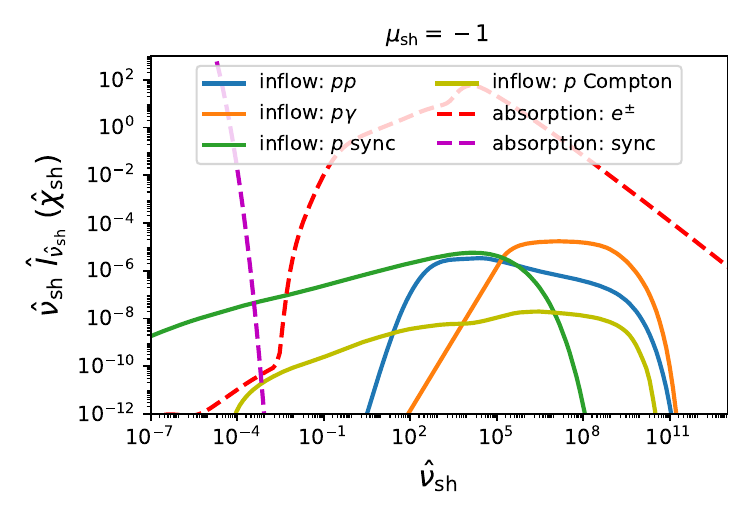}
\caption{Photon spectrum in the immediate downstream region ($\hat{\tau} = 0.1$), obtained by accounting for both leptonic and hadronic interactions, with $\sigma_u = 0.1$. {\it Top and middle:}  Spectra of downstream-going and upstream-going photons, respectively. The black and red solids lines represent the spectral intensities for the configurations  with and without hadronic processes. {\it Bottom:} Contributions from $pp$, $p\gamma$, proton synchrotron, and proton Compton scattering to the inflow from the downstream boundary (solid lines) and leptonic absorption mechanisms (pair production and annihilation as well as synchrotron) that significantly absorb the photons emitted by non-thermal particles (dashed lines).}
\label{fig:had_spectrum}
\end{figure}

Figure~\ref{fig:had_spectrum} shows the resultant photon spectral energy distribution at $\hat{\tau} = 0.1$. In the presence of hadronic processes, the high-energy photons from the far downstream region modify the photon spectrum at the subshock. As visible  from the bottom panel, in the high-energy spectral tail (i.e., for $\hat{\nu}_{\rm sh} > 10^5$), the emission is dominated by $p\gamma$ inelastic collisions. However, in the low-energy spectral tail (i.e., for $10^2 < \hat{\nu}_{\rm sh} < 10^5$), the emission is dominated by $pp$ interactions. At smaller energies (i.e., for $\hat{\nu}_{\rm sh} < 10^2$),  proton synchrotron radiation is the main  hadronic emission channel. However, its intensity is several orders of magnitude lower than that of  leptonic processes and hence negligible. 

As shown in Fig.~\ref{fig:had_spectrum} (bottom panel), at the subshock, the upstream-going photon spectrum is not identical to the incoming photon spectrum at the downstream boundary. This is because the photons produced by hadronic processes are absorbed. For example, for $10 < \hat{\nu}_{\rm sh} < 10^6$, strong absorption caused by pair production occurs, resulting in a dip between the leptonic and hadronic peaks. At very low frequencies, synchrotron self-absorption dominates, so photons produced by proton synchrotron radiation cannot survive.

\section{Conclusions and outlook}
\label{sec:conclusions}

Radiation mediated shocks form when a shock propagates in an optically thick plasma and are common in astrophysical transients, such as supernovae, neutron-star mergers, and gamma-ray bursts. In order to interpret the multi-messenger emission from these sources, it is important to understand the feedback of leptonic and hadronic interactions on the shock physics. For this purpose, for the first time, we investigate the physics of a steady-state, radiation-mediated, planar  shock, accounting for  electron and proton acceleration processes, as well as variable upstream magnetization ($\sigma_u = 0$, $10^{-8}$, $10^{-4}$, $0.1$, and $0.3$) and fixed relativistic upstream bulk fluid velocity ($\Gamma_u = 10$). 

For non-magnetized shocks and in the absence of proton acceleration, Compton scattering together with pair production and annihilation dominate the photon emission and absorption, in agreement with the findings of Ref.~\cite{2010ApJ...725...63B}. The downstream plasma eventually reaches equilibrium with the radiation, and the upstream-going photons decelerate the inflowing material before it reaches the shock plane, making the shock profile smoother. 

For magnetized shocks, electrons in the immediate downstream can be heated to  relativistic temperatures  ($T \simeq 1~{\rm MeV}$), enhancing synchrotron emission. Although for $\sigma_u \leq 10^{-4}$, such electron synchrotron radiation is negligible, synchrotron self-absorption is important even at low magnetization (i.e., $\sigma_u \simeq 10^{-8}$).  Specifically, synchrotron self-absorption dominates  the low-energy tail of the photon spectral energy distribution in the shock upstream (photons scattered into this energy range by electrons in the cold upstream are absorbed and cannot  propagate further);  the downstream-going photon spectral energy distribution is also flattened. As a consequence,  the spectral distribution of downstream-going photons and the shock dynamics in the upstream region are altered. With moderately high magnetization (i.e., $\sigma_u \gtrsim 0.1$), synchrotron radiation becomes more prominent, flattening the  upstream-going photon spectrum. In general, due to the deceleration of the upstream fluid, the shock jump becomes sharper as $\sigma_u$ increases, with a prominent subshock forming for $\sigma_u \gtrsim 0.1$, in agreement with the findings of  Ref.~\cite{2017ApJ...838..125B}.

Accounting for proton acceleration, a high-energy spectral component arises in the photon spectrum in the immediate downstream because of the pile-up of high-energy photons,    when protons can be efficiently accelerated at the subshock for $\sigma_u \gtrsim 0.1$. However, the contribution of  the high-energy photon component  to the radiation flux and pressure is negligible, and thus  the shock dynamics is negligibly affected. We note that,  in  neutron-rich environments, such as short gamma-ray bursts, proton–neutron and neutron–neutron collisions may further affect the shock structure. In this case, neutrons can propagate both upstream and downstream, unaffected by collective plasma effects~\cite{2017ApJ...838..125B}.

We do not investigate shock configurations with  $\sigma_u \gtrsim 1$, because higher magnetization leads to a faster electron cooling, and the required resolution would be too high to achieve at present. However, for $\sigma_u = 0.3$, the velocity jump at the subshock is already nearly half of the initial jump in the absence of radiation, allowing us to explore the formation of a subshock at which particles can be accelerated efficiently. For larger $\sigma_u$, we expect  that subshock should be more prominent. 

Future work should also investigate the shock physics for different $\Gamma_u$,  leading to varying amounts of internal energy  in the downstream. Although pair production and annihilation may keep the plasma temperature approximately constant, the radiation intensity and peak frequency can still be altered, especially in the presence of a strong magnetic field. 

We  focused on exploring the impact of electron and proton acceleration on the photon spectral energy distribution. However, because of the feedback of the latter on the shock properties, high-energy neutrinos may also be produced~\cite{2013PhRvL.111l1102M,2016PhRvD..93e3010T}. The shape of the neutrino distribution and its characteristic energy are expected to depend on the efficiency of particle acceleration at the subshock and the jet baryon loading below the photosphere~\cite{2023PhRvD.107b3001G,2024ApJ...961L...7R,2025ApJ...983...34R}. Although high-energy neutrino emission does not directly impact  the  shock structure,  variations of the shock structure in the radiation-dominated regime are expected to influence the neutrino signal. 

Our work highlights the importance of modeling the shock dynamics accounting for the feedback of the radiation produced in the aftermath of particle acceleration. For simplicity, we focused on a steady-state shock. Accounting for the temporal evolution of shock formation and pair production could drive the system into a nonlinear phase (e.g., due to pair production runaway~\cite{2001A&A...374..719P}), such that the stationary solution no longer holds. However, our findings highlight that further work is needed to better understand shock-radiation feedback mechanisms and reliably forecast the multi-messenger signals from astrophysical transients.

\begin{acknowledgments}
We thank Ore Gottlieb and Eli Waxman for valuable discussions. This project has received support from the Villum Foundation (Project No.~13164) and the  German Research Foundation (DFG) through the Collaborative Research Centre ``Neutrinos and Dark Matter in Astro- and Particle Physics (NDM),'' Grant No.~SFB-1258-283604770. 
The Tycho supercomputer hosted at the SCIENCE HPC Center at the University of Copenhagen was used to support the numerical simulations presented in this work.
\end{acknowledgments}

\appendix

\section{\label{app:leptonic}Leptonic processes}

In this appendix, we outline our modeling of Compton processes, pair production and annihilation, Bremsstrahlung, as well as synchtrotron and its self-absorption. All these processes are of relevance to our forecast of the photon spectral energy distribution. 

\subsection{Compton scattering}

The total cross section for Compton scattering is given by~\cite{1979rpa..book.....R}
\begin{eqnarray}
\sigma_c (\hat{\nu},\hat{T}) &=& \sigma_T \frac{3}{4}\Bigg[\frac{1+\xi}{\xi^3}\left\{\frac{2\xi(1+\xi)}{1+2\xi} - \ln(1+2\xi)\right\} \nonumber \\ 
&&+ \frac{\ln(1+2\xi)}{2\xi} - \frac{1+3\xi}{(1+2\xi)^2}\Bigg]\, ,
\end{eqnarray}
where $\xi = \hat{\nu}(1+2\hat{T})$ incorporates a correction due to the plasma temperature. 

The differential cross section is
\begin{eqnarray}
\label{eq:Compton}
    \frac{d\tilde{\sigma_s}}{d\hat{\nu}^{\prime}}(\hat{\nu}^{\prime}\rightarrow \hat{\nu}) = \frac{1}{4\pi}\sigma_c(\hat{\nu}^{\prime},\hat{T}) f_d(\hat{\nu}^{\prime},\hat{T},\hat{\nu})\, ,
\end{eqnarray}
where $\hat{\nu}^{\prime}$ and $\hat{\nu}$ represent the frequencies of the seed photon and the scattered photon, respectively. The scattered photons are assumed to be isotropic in the rest frame of the target electrons. The function $f_d(\hat{\nu}^{\prime}, \hat{T}, \hat{\nu})$ denotes the photon redistribution function (i.e.,  the probability density for a seed photon with frequency $\hat{\nu}^{\prime}$ to be scattered to a frequency $\hat{\nu}$) and  is calculated following  Appendix B of Ref.~\cite{2010ApJ...725...63B}.

When $\hat{T} \leq 0.25$, 
\begin{eqnarray}
    f_d(\hat{\nu}^{\prime}, \hat{T}, \hat{\nu}) = A \hat{\nu}^3e^{-\hat{\nu}/T_{\rm eff}}\, ,
\end{eqnarray}
with $A = 1 / (6\hat{T}_{\rm eff}^4)$ and $T_{\rm eff} = \hat{\nu}_0(\hat{\nu}^{\prime},\hat{T})/4$. The photon spectrum  after Compton scattering is approximately a thermal spectrum with an effective temperature $T_{\rm eff}$ and an average frequency $\hat{\nu}_0$, which can be estimated as:
\begin{eqnarray}
    \frac{\hat{\nu}_0}{\hat{\nu}^{\prime}} =  {\rm min}\left[\left( 1 + \frac{4\hat{T}(4\hat{T}+1)-\hat{\nu}^{\prime}(\hat{\nu}^{\prime}+1)}{(1+a_{\nu}(\hat{T}))\hat{\nu}^{\prime}} \right), \frac{4\hat{T}}{\hat{\nu}^{\prime}}\right] \nonumber \\
\end{eqnarray}
for $\hat{\nu}^{\prime} \leq 4\hat{T}$ and
\begin{eqnarray}
    \frac{\hat{\nu}_0}{\hat{\nu}^{\prime}} = \frac{1}{1 + {\rm ln}\left[{(\hat{\nu}^{\prime}+1)}/{(4\hat{T}+1)}\right]}
\end{eqnarray}
for $\hat{\nu}^{\prime} > 4\hat{T}$. The function  $a_{\nu}$ is defined as
\begin{eqnarray}
    a_{\nu}(\hat{T}) = &-&0.003763{\rm ln}^4(\hat{T})-0.0231 {\rm ln}^3(\hat{T})  \\
    &-& 0.01922 {\rm ln}^2(\hat{T}) - 0.129 {\rm ln}(\hat{T}) + 3.139\, ,\nonumber 
\end{eqnarray}
with $a_{\nu}(\hat{T} < 0.01) = a_{\nu}(\hat{T} = 0.01)$. 

For $\hat{T} > 0.25$, in the Klein-Nishina regime with $\hat{\nu}^{\prime} > 1/(4\hat{T})$, we have
\begin{eqnarray}
    f_d(\hat{\nu}^{\prime}, \hat{T}, \hat{\nu}) = \frac{\hat{\nu}^2 e^{-\hat{\nu}/\hat{T}}\sigma_c(\hat{\nu}^{\prime},\hat{T})}{a_d(\hat{T})\hat{T}^3\sigma_T}\, ,
\end{eqnarray}
where
\begin{eqnarray}
    a_d(\hat{T}) = &-&0.0046 {\rm ln}^4(\hat{T}) + 0.007197{\rm ln}^3(\hat{T})  \\ 
    &+& 0.09079 {\rm ln}^2(\hat{T}) - 0.3166{\rm ln}(\hat{T}) + 0.3146\, , \nonumber 
\end{eqnarray}
and $a_d(\hat{T}) = a_d(5)(\hat{T}/5)^{-1.7}$ is set for $\hat{T} > 5$. In the inverse Compton regime, we have
\begin{eqnarray}
    f_d(\hat{\nu}^{\prime},\hat{T}, \hat{\nu}) \propto \sqrt{\hat{\nu}}e^{-\sqrt{\frac{\hat{\nu}}{\frac{4}{3}\hat{T}^2 \hat{\nu}^{\prime}}}}\Theta(\hat{\nu} - \hat{\nu}^{\prime})\, ,
\end{eqnarray}
where $\Theta(\hat{\nu} - \hat{\nu}^{\prime})$ is a step function, which has $\Theta(\hat{\nu} - \hat{\nu}^{\prime}) = 0$ when $\hat{\nu} < \hat{\nu}^{\prime}$, and $\Theta(\hat{\nu} - \hat{\nu}^{\prime}) = 1$ when $\hat{\nu} > \hat{\nu}^{\prime}$. The distribution function is normalized such that  $\int_0^{+\infty} f_d(\hat{\nu}^{\prime}, \hat{T}, \hat{\nu}) d\hat{\nu} = 1$.

\subsection{Pair production and annihilation}
For a given  plasma temperature $\hat{T}$,  the $e^{\pm}$  annihilation rate is~\cite{1982ApJ...258..335S}
\begin{eqnarray}
    r_{e^{\pm}}(\hat{T}) &=& \frac{3}{4}\left[1 + \frac{2\hat{T}^2}{\ln\left(2\eta_E \hat{T} + 1.3\right)}\right]^{-1}\, ,
\end{eqnarray}
where $\eta_E = 0.5616$. The spectral distribution of the resultant photons can be written as~\cite{1982ApJ...258..335S}
\begin{eqnarray}
f(\hat{\nu},T) \propto&& \frac{1}{\hat{\nu}}
{\rm exp}[x_1 T_7^{-0.12} + x_2 T_7^{-0.16}{\rm ln \hat{\nu}} + x_3 T_7^{-0.52}({\rm ln} \hat{\nu})^2 \nonumber\\
&&+ x_4 T_7^{-1.84}({\rm ln}\hat{\nu})^3 + x_5 T_7^{-1.94}\hat{\nu}]
\end{eqnarray}
where $T_7 = T /(10^7$~K$)$, and $x_i = \sum\limits_{j=1}^5 a_{ij}\left({\rm ln T_7}\right)^{j-1}$. The coefficients $a_{ij}$ are provided in Table 1 of Ref.~\cite{1982ApJ...258..335S}, and $x_4$ is set to zero for $T > 4\times 10^{10}$~K.

The cross section for $\gamma \gamma$ collisions producing $e^{\pm}$ pairs is given by~\cite{1979rpa..book.....R}
\begin{eqnarray}
\sigma_{\gamma \gamma} (s) &=& \frac{3}{8}\,\frac{\sigma_T}{s} \Bigg[ \left(2 + \frac{2}{s} - \frac{1}{s^2}\right)\cosh^{-1}\!\left(s^{1/2}\right) \nonumber \\
&&- \left(1+\frac{1}{s}\right)\left(1-\frac{1}{s}\right)^{1/2} \Bigg]\, ,
\end{eqnarray}
where
\begin{eqnarray}
s &=& \frac{1}{2}\, h\nu \, h\nu^{\prime} (1 - \mu \mu^{\prime})
\end{eqnarray}
combines the frequency and directional information of the  incoming photons.

\subsection{Bremsstrahlung}
The numerical factor $\lambda_{\rm ff}$ in Eqs.~\ref{eq:eta_ff} and \ref{eq:chi_ff} is given by
\begin{eqnarray}
\lambda_{\rm ff}(a_{e^{\pm}}, T) &=& (1+a_{e^{\pm}})\lambda_{ep} + [a_{e^{\pm}}^2 + (1+a_{e^{\pm}})^2]\lambda_{ee} \nonumber \\ &&+ a_{e^{\pm}} (1+a_{e^{\pm}})\lambda_{e^{\pm}}
\end{eqnarray}
where $\lambda_{ep}$, $\lambda_{ee}$, and $\lambda_{e^{\pm}}$ are the numerical factors corresponding to $e^{\pm}p$, $e^{\pm}e^{\pm}$, and $e^{+}e^{-}$ thermal bremsstrahlung, respectively. For each case, the relation $\lambda = \pi/(2\sqrt{3})g$ holds, and the analytical expressions for $g_{ep}$, $g_{ee}$, and $g_{e^{\pm}}$ are provided in the appendix of Ref.~\cite{1995ApJ...446...86S} (note that an erratum letter was published for some of the equations in the original paper).

\subsection{Synchrotron radiation and self-absorption}
Considering an electron (positron) with energy $E_e$, gyrating around a magnetic field with strength $B$, its  radiation power can be calculated 
as~\cite{1979rpa..book.....R}
\begin{eqnarray}
P_{\nu,{\rm syn}}(\nu,E) = \frac{\sqrt{3}e^3B}{m_e c^2} F\left(\frac{\nu}{\nu_c}\right)\, ,
\end{eqnarray}
with 
\begin{eqnarray}
F(x) = x\int_x^{\infty}K_{5/3}(\xi)d\xi\, ,
\end{eqnarray}
and $K_{5/3}(\xi)$ is the modified Bessel function of second kind. 

We assume that the random-motion Lorentz factor of the relativistic thermal electrons (positrons) is described by a Maxwell–J$\ddot{\text{u}}$ttner distribution with a characteristic temperature, which reads as~\cite{1911AnP...339..856J}
\begin{eqnarray}
f_{e^{\pm}}(\gamma_e, \hat{T}) = \frac{\gamma_e^2 \beta_e}{\hat{T} K_2(1/\hat{T})}e^{-\gamma_e/\hat{T}}\, ,
\end{eqnarray}
where $K_2(1/\hat{T})$ is the modified Bessel function of the second kind. The energy distribution of thermal electrons (positrons), $f_{e^{\pm}}(E_e)$, can be obtained from $f_{e^{\pm}}(\gamma_e)$, exploiting the fact that $E_e = \gamma_e m_e c^2$.

\subsection{\label{sec:lep_contribution}Impact of leptonic processes}
\begin{figure*}
\centering
\includegraphics[width = 1.98\columnwidth]{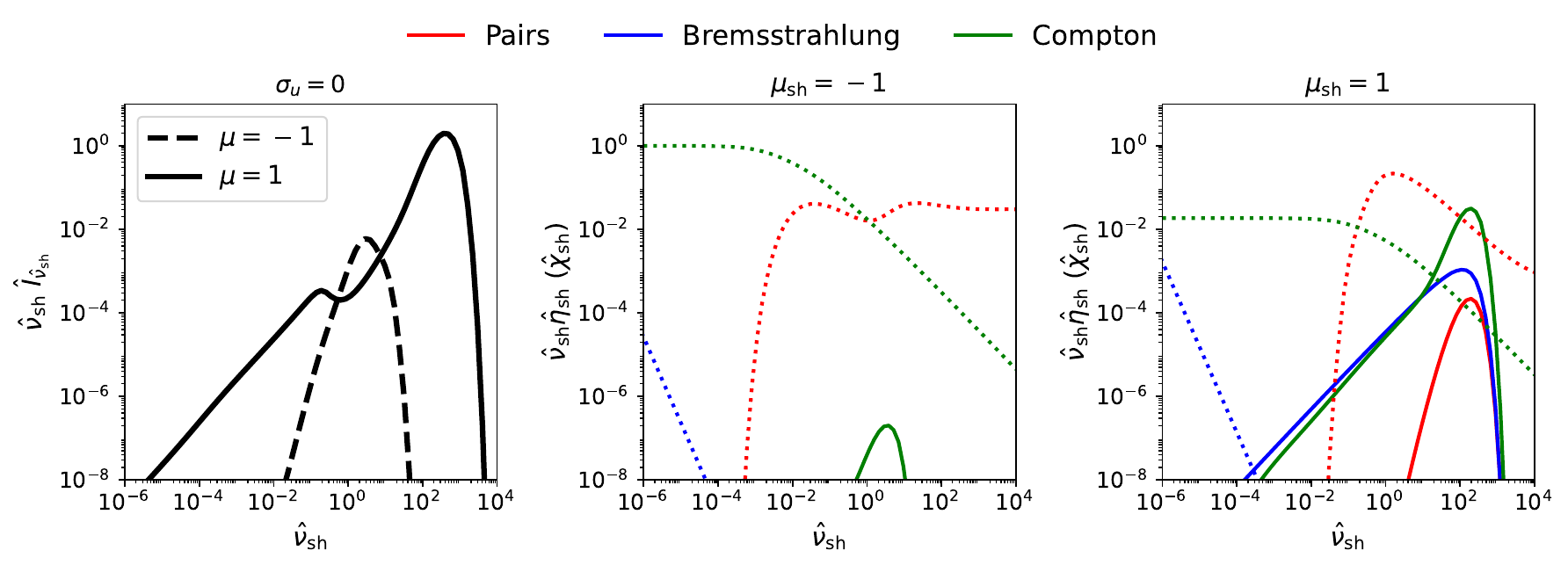}
\caption{Photon spectrum as well as emission and absorption coefficients for a radiation-mediated shock with $\Gamma_u = 10$ and $\sigma_u = 0$ at $\hat{\tau} = -20$. {\it Left}: Upstream- and downstream-going ($\mu_{\rm sh} = -1$ and $\mu_{\rm sh} = 1$) photon spectra are shown with dashed and solid lines, respectively. {\it Middle:} Emission and absorption coefficients for the upstream-going photons are plotted  with solid and dotted lines, respectively.  The red, blue, and green lines represent the coefficients for pair processes, bremsstrahlung, and Compton scattering. {\it Right:} Same as the middle panel, but for the downstream-going photons. }
\label{fig:eta_chi_sigma0}.
\end{figure*}

\begin{figure*}
\centering
\includegraphics[width = 1.98\columnwidth]{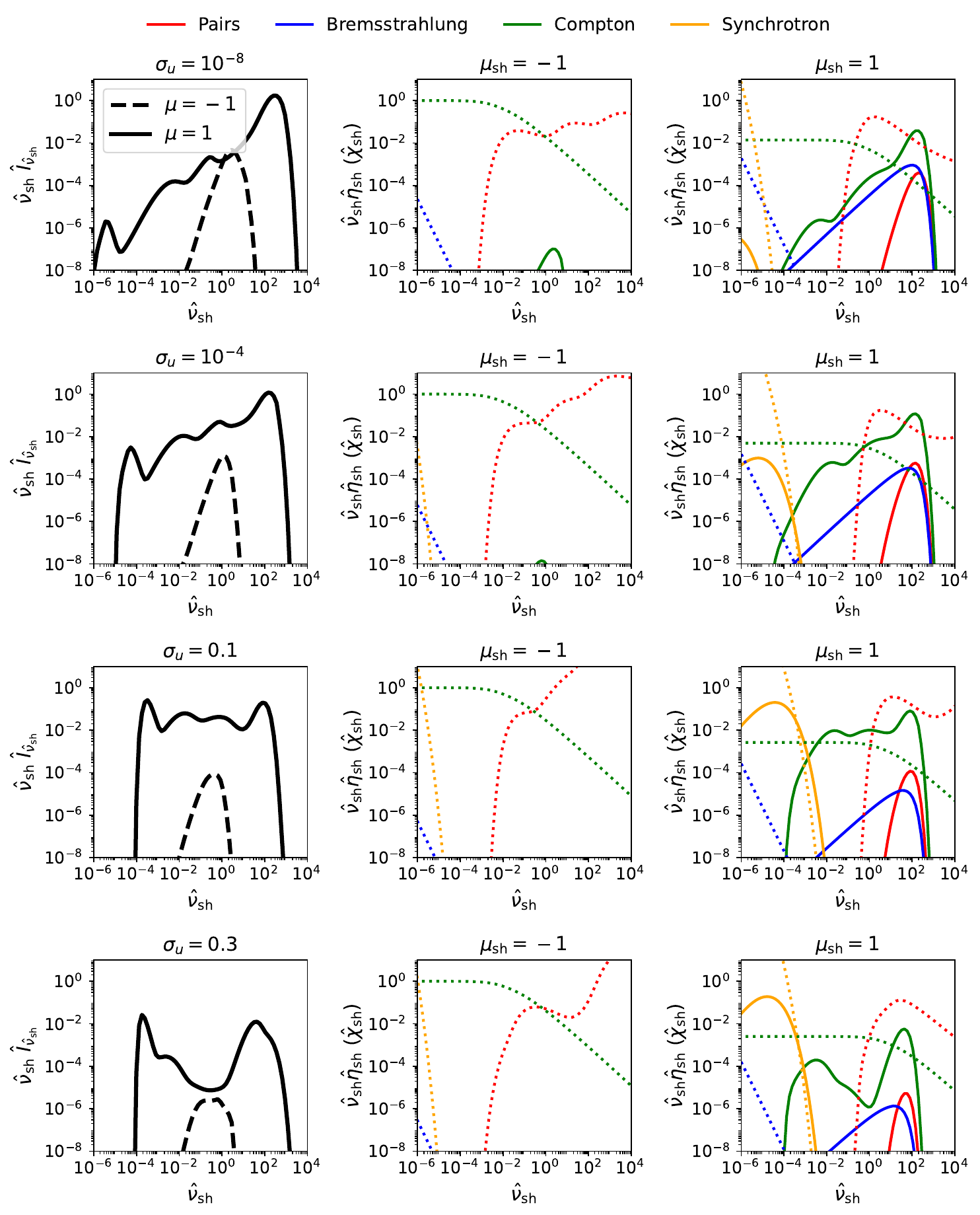}
\caption{Similar to Fig.~\ref{fig:eta_chi_sigma0}, but for magnetized radiation-mediated shocks with $\sigma_u = 10^{-8}$, $10^{-4}$, $0.1$, and $0.3$ from the top to bottom panels. At $\hat{\tau} = -20$, the case with high magnetization has a lower photon intensity, indicating that photons are absorbed more efficiently. Synchrotron self-absorption dominates over other absorption mechanisms for low-energy photons and affects  the downstream-going photons more prominently than on the upstream-going photons.}
\label{fig:eta_chi}
\end{figure*}

Figures~\ref{fig:eta_chi_sigma0} and \ref{fig:eta_chi} show the photon spectrum in the upstream (at $\hat{\tau} = -20$) for $\sigma_u=0$ and $\sigma_u=10^{-8}$, $10^{-4}$, $0.1$ as well as $0.3$. 

In a non-magnetized or weakly magnetized shock, emission and absorption are dominated by Compton scattering together with pair production and annihilation, both upstream and downstream. The bremsstrahlung absorption coefficient increases towards the low-energy tail of the photon spectrum, dominating the absorption at low frequencies for $\sigma_u=0$. Synchrotron losses  become significant  for $\sigma_u \gtrsim 10^{-8}$. Synchrotron self-absorption does not affect the upstream-going photons. However, once the photons are Compton scattered to lower energies by cold electrons in the upstream, they can be absorbed  efficiently. Since the Compton-scattered photons are boosted by the relativistic bulk motion of the upstream fluid, they appear to travel towards $\mu_{\rm sh} = 1$ in the shock frame. Consequently, synchrotron self-absorption is more pronounced for the downstream-going photons ($\mu_{\rm sh} = 1$). For larger $\sigma_u$,  the photon intensity is lower. This indicates that upstream-going photons are efficiently absorbed  in a magnetized fluid; hence, the characteristic physical quantities at this position are closer to their far-upstream ones, consistent with  Fig.~\ref{fig:lep_profile}.

\section{Proton acceleration and cooling}
\label{app:acc_cool}
The acceleration rate of protons is provided in Eq.~\ref{eq:p_acc}. Several cooling mechanisms constrain the maximum energy of accelerated protons. Any physical quantity not explicitly associated with a specific reference frame in this section is defined in the comoving frame of the  downstream fluid. 

The cooling rate due to inelastic $pp$ collisions is~\cite{1997PhRvL..78.2292W}
\begin{eqnarray}
t_{\rm pp}^{-1} = c n_p \sigma_{pp}\, ,
\end{eqnarray}
where $\sigma_{pp} \approx 3.0\times 10^{-26}~{\rm cm}^2$ is the cross section of $pp$ collisions; $n_p$ is the number density of background protons. The cooling rate due to inelastic $p\gamma$ collisions is calculated as
\begin{eqnarray}
t_{p\gamma}^{-1}
&= \frac{c}{2 \gamma_{p}^{2}} \int_{\epsilon_{\text{th}}}^{\infty} 
d\bar{\epsilon}_{\gamma} \, \sigma_{p\gamma}\left(\bar{\epsilon}_{\gamma}\right) 
\kappa_{p\gamma}\left(\bar{\epsilon}_{\gamma}\right)  \bar{\epsilon}_{\gamma}
\int_{\bar{\epsilon}_{\gamma}/(2\gamma_{p})}^{\infty} 
\frac{d\epsilon_{\gamma}}{\epsilon_{\gamma}^{2}} n_{\epsilon_{\gamma}}\ , \nonumber \\
\end{eqnarray}
where $n_{\epsilon_{\gamma}} = {[2\pi \int I^{\prime}(\mu^{\prime},\nu^{\prime})d\mu^{\prime}]}/{[hc\nu]}$ is the specific number density of photons at an energy $\epsilon_{\gamma}$, with $\epsilon_{\gamma} = h \nu$.  The photon energy in the proton rest frame is $\bar{\epsilon}_{\gamma}$. We adopt the fit to the inelastic $p\gamma$ cross section data provided in the appendix of Ref.~\cite{2008MNRAS.385.1461Y}. The coefficient $\kappa_{p\gamma} = 0.2$ is adopted as the inelasticity parameter, and $\epsilon_{\rm th} = 0.15~{\rm GeV}$ is the threshold photon energy for $p\gamma$ collisions. 

The cooling rate due to proton synchrotron radiation is calculated as~\cite{1979rpa..book.....R}
\begin{eqnarray}
t_{\rm p,syn}^{-1} =  \frac{2\sigma_T m_e^2 c}{3 m_p^2} \gamma_p n_p \sigma\ ,
\end{eqnarray}
where $\sigma$ denotes the local magnetization parameter, as defined in Eq.~\ref{eq:B_flux_cons}.

\begin{figure}
\includegraphics[width = 0.99\columnwidth]{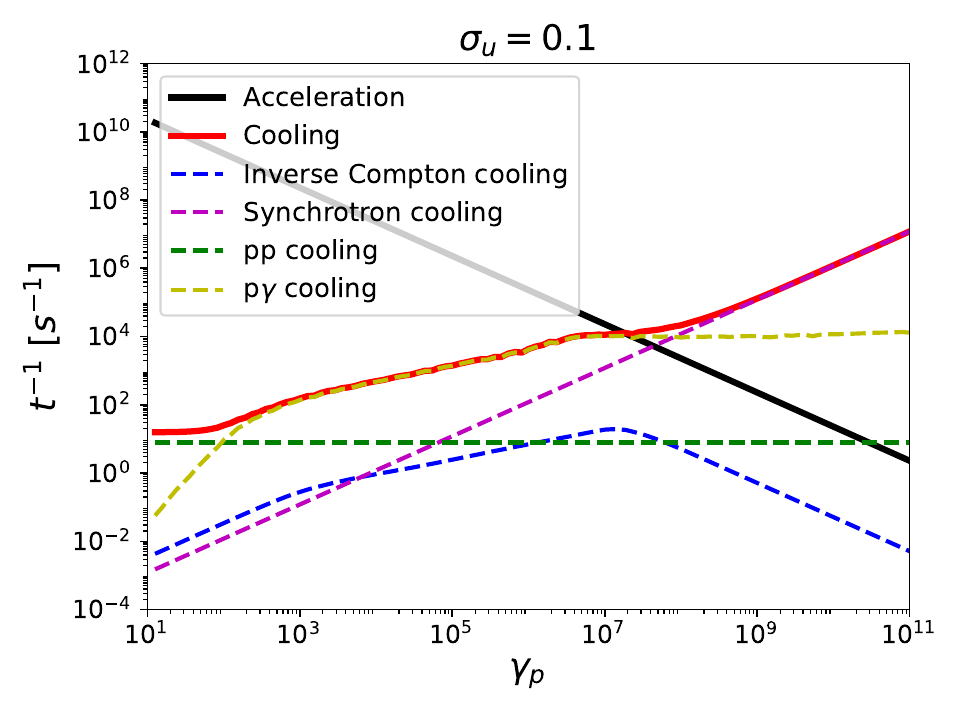}
\label{fig:p_acc_cooling}
\caption{Acceleration and cooling rates of protons   in the immediate downstream  and for $\sigma_u = 0.1$. The solid black and red lines represent the acceleration  and the  cooling rates as functions of the proton  Lorentz factor. The dashed lines indicate the cooling rates for inverse Compton, synchrotron, $pp$ and $p\gamma$ interactions, in blue, magenta, green, and yellow, respectively. 
}
\end{figure}

The cooling rate of inverse Compton scattering between protons and photons can be described by a piecewise function~\cite{2003PhRvD..68h3001R}:
\begin{eqnarray}
    t_{\rm IC}^{-1} = \left\{
    \begin{array}{cc}
      \frac{4\sigma_T m_e^2 E_p \epsilon_{\gamma} n_{\epsilon_\gamma}}{3m_p^4 c^3},   & ~~~~E_p \epsilon_{\gamma} \leq m_p^2 c^4\, , \\
      \frac{4\sigma_T m_e^2 c^5 n_{\epsilon_\gamma}}{3E_p \epsilon_{\gamma}},   & ~~~~E_p \epsilon_{\gamma} < m_p^2 c^4\, .
    \end{array}
    \right. 
\end{eqnarray}
The total cooling rate is calculated as $t^{-1}_{\rm cooling} = t_{\rm pp}^{ -1} + t_{p\gamma}^{-1} + t_{\rm p,syn}^{-1} + t_{\rm IC}^{-1}$. To estimate the maximum energy that a proton can achieve, one should compare the acceleration rate to the total cooling rate. The acceleration and cooling rates as a function of the proton Lorentz factor is shown in Fig.~\ref{fig:p_acc_cooling} for $\sigma_u = 0.1$. 

%


\end{document}